\documentclass[12pt,apj]{emulateapj}

\usepackage{subfigure}
\usepackage[usenames]{color}
\usepackage{enumerate}
\usepackage{soul}

\begin{document}
\shorttitle{Multi-Parametric Study of Rising Buoyant Flux Tubes}
\shortauthors{Mart\'inez-Sykora, Moreno-Insertis \& Cheung }

\title{Multi-Parametric Study of Rising 3D Buoyant Flux Tubes in an Adiabatic Stratification Using AMR} 

\author{Juan Mart\'inez-Sykora$^{1,2}$}
\author{Fernando Moreno-Insertis$^{3,4}$}
\author{Mark C. M. Cheung$^{1}$}

\affil{$^1$ Lockheed Martin Solar and Astrophysics Laboratory, Palo Alto, CA 94304, USA}
\affil{$^2$ Bay Area Environmental Research Institute, Petaluma, CA, USA}
\affil{$^3$ Instituto de Astrof\'isica de Canarias, 38200 La Laguna (Tenerife), Spain}
\affil{$^4$ Department of Astrophysics, Faculty of Physics, University of La Laguna, 38200 La Laguna (Tenerife), Spain}

\newcommand{\komment}[1]{\texttt{#1}}
\newcommand{\mcmc}[1]{MC: \textbf{#1}}
\newcommand{\blank}[1]{{\em \textbf{#1}}}
\newcommand{\Fret}{{\Phi_{\rm ret}}}
\newcommand{\Beq}{{B_{eq}}}
\newcommand{\chimone}{{\chi^{-1}_{}}}

\begin{abstract}

We study the buoyant rise of magnetic flux tubes embedded in an adiabatic
stratification using two-and three-dimensional, MHD
simulations. We analyze the dependence of the tube evolution on the field line
twist and on the curvature of the tube axis in different diffusion regimes. 
To be able to achieve a comparatively high spatial resolution we use the
FLASH code, which has a built-in Adaptive Mesh Refinement (AMR)
capability. Our 3D experiments reach
Reynolds numbers that permit a reasonable comparison of the results with those
of previous 2D simulations. When the experiments are run without AMR, 
hence with a comparatively large diffusivity, the amount of
longitudinal magnetic flux retained inside the tube increases with the
curvature of the tube axis. However, when a 
low-diffusion regime is reached by using the AMR algorithms, the magnetic
twist is able to prevent the splitting of the magnetic loop into
vortex tubes and the loop curvature does not play any significant role.
We detect the generation of
vorticity in the main body of the tube of opposite sign on the opposite sides
of the apex. This is a consequence of the inhomogeneity of the azimuthal
component of the field on the flux surfaces. The lift force associated with this global
vorticity makes the flanks of the tube move away from their initial vertical
plane in an antisymmetric fashion. The trajectories have an
oscillatory motion superimposed, due to the shedding of vortex rolls
to the wake, which creates a Von Karman street.  \end{abstract}


\maketitle

\section{Introduction}

Starting with the seminal paper of \citet{Parker:1977sf}, a large body of
work along the past four decades has shed light on the key
magnetohydrodynamic (MHD) processes that affect the rise of magnetic flux
tubes across a stratified medium like the solar convection zone. An 
approach, not followed in this paper,
is to study the large-scale flux tube evolution by way of simple
one-dimensional models using the {\em thin magnetic flux tube approximation}
\citep{spruit1981,Roberts1978}: such models allowed the study of the
large-scale motion of flux tubes from the bottom of the convection zone to
the near-surface layers \citep[e.g.,][]{moreno1986,
  DSilva:1993zr,fan1993,Fan:1994ly, Caligari1995,Weber:2011fk}, including a
detailed stratification of the solar interior. 
In those papers, the moving magnetic tube and the stratified
background are assumed to be distinct  media, with the latter influencing, but
being unperturbed by, the former. The thin flux tube approximation 
assumes the radius of the flux tube to be much smaller
than the other relevant length-scales in the problem (including pressure scale height and
radius of curvature of the tube) and does not allow for the study of how the tube
cross-section evolves. A new class of models that study the 3D
evolution of magnetic tube-like concentrations rising through the depth range
of the convection zone has appeared in recent years \citep{Fan:2008qf,
Jouve:2009bh,Jouve:2013kx,Pinto:2013uq}: in those papers, the anelastic
three-dimensional fluid-dynamics equations are solved for a spherical shell
with the stratification of the solar convection zone. The last three of those
models include self-consistent large-scale convective motions that encompass
the whole range of depths of the simulated domain. Those calculations
are computationally demanding so they cannot reach high viscous and magnetic
Reynolds numbers. On the other hand, it is interesting to note that, in what
concerns the motion of the tube-like domains, they recover many of the basic
results obtained with the older thin-tube models.

A different approach followed in the past decades, and also in the present
paper, is to study the internal dynamics of buoyant magnetic tube-like
regions by using two- or three-dimensional models with a simplified,
adiabatic and cartesian background stratification. The simulations of
\citet{Shussler1979,longcope1996, moreno1996,emonet1998,Fan:1998vn,
Hughes:1998ty, emonet2001} and \citet{mark2006} among others, in 2D, and 
\citet{Dorch:1998db}, \citet{Dorch:1999fk}, \cite{Fan:1999uq} 
and \citet{abbett2000}, in 3D, examined, in particular, the
physical mechanisms whereby the moving magnetic flux tubes can maintain (or
otherwise lose) their coherence during the rise.
\citet{Shussler1979} showed that an untwisted, rising horizontal tube 
splits into counter-rotating vortices after rising a height comparable to the
initial tube diameter. \citet{longcope1996} carried out a similar experiment
with higher spatial resolution; they explained the sideways motion 
of the counter-rotating vortex tubes through the combined
action of the buoyancy and lift forces exerted by the surrounding
flow. \citet{moreno1996} and \citet{emonet1998} demonstrated that a
sufficient amount of twist of the magnetic field lines can
provide the magnetic tension necessary to prevent the splitting of the
tube. They obtained a criterion, namely, $\tan \Psi \gtrsim (R/H_p)^{1/2}$,
with $R$ the tube radius, $H_p$ the pressure scale-height, and $\Psi$ the
pitch angle of the field lines in the dynamically relevant part of the tube,
for the splitting into sideways expanding vortices to be substantially
reduced, with the tube thus retaining a good fraction of its original
magnetic flux. Additionally, \citet{emonet1998}
explained the importance of the magnetic boundary layer at the periphery of
the tube as a site of vorticity generation: vorticity-loaded fluid elements
sliding along the boundary layer are carried to the wake where they
accumulate. This is a process reminiscent of the behavior of flows around rising air
bubbles in liquids or past rigid cylindrical rods 
\citep[e.g.,][]{daviestaylor50, collins65, parlange69,Berger:1972vn,
  wegenerparlange73, hnatbuck76, ryskinleal84b, christovvolkov85, 
Williamson:1988fk,Williamson:1988uq,Norberg:1994ys,
Williamson:1996kx,Rast:1998zr}.

The phenomenon of vortex shedding from the wake was analyzed by
\citet{Fan:1998vn} and \citet{emonet2001}: the vortex rolls can be ejected
alternatively from each side of the main tube creating a trailing Von
K\'arm\'an vortex street: each emission of a vortex roll leaves the head with
an excess vorticity of the same amount and opposite sign. The alternation in
the sign of the vorticity of the head entails an oscillating lift force
which, when combined with the drag and buoyancy forces, leads in the simplest
cases to an oscillatory sideways motion of the tube's head superimposed on
its rise. As shown by \citet{emonet2001}, in more complicated situations,
when the amount of vorticity in the wake is large, vortex shedding leads to
large lift forces in the tube head; when the lift force is dominant, the
complication of the trajectory of the tube head can be considerable (see,
e.g., their Figure~6). This is probably the reason for the peculiar trajectory
resulting from the simulation by \citet{Hughes:1998ty}, who used an
adaptive-grid numerical code that allowed them to reach high Reynolds
numbers.

The importance of reaching high Reynolds numbers in this kind of experiment
was clearly demonstrated in the 2D simulations by
\citet{mark2006}. These authors used the FLASH code \citep{Fryxell:2000rr},
and applied the code's Adaptive Mesh Refinement (AMR) feature to study in
detail the dependence of the results on the effective Reynolds number of the
flows; the latter was determined by finding the ratio of the actual thickness
of the boundary layer measured in the experiment to the tube diameter. When
the magnetic Reynolds number ($Re_m$) reaches values around several 
hundred, the smooth vorticity profile of the wake characteristic of 
lower-$Re_m$ experiments turns into a complicated
distribution with a hierarchy of vortices down to very small
sizes. Importantly, they also showed that the fraction of magnetic flux
retained inside the tube greatly increases when $Re_m$ grows from below $100$
up to a few hundred; the increase seems to level off when $Re_m$ enters the
range around a few thousand. In addition, the authors confirmed 
previous results in that the amount of flux retained in the tube's head
decreases with decreasing levels of field line twist around the axis even in
the high-resolution, high-$Re_m$ regime. They also reproduced the
vortex-shedding phenomenon observed in lower-$Re_m$ calculations.

There are also a few instances of three-dimensional simulations that study
the coherence of magnetic tubes rising in idealized
stratifications. \citet{Dorch:1998db} studied the effect of the transverse
component of the magnetic field on the tube's evolution: their tubes with a
purely axial (i.e., untwisted) field suffered the splitting into vortex tubes
also obtained by the authors reported above. In a few of their
experiments, however, they introduced a transverse magnetic field
superimposed on the axial component: the transverse field was a mixture of a
simple twisted structure (obtained via the curl of a Gaussian-shaped vector
potential) and a random component. The resulting strength of the transverse
component was high enough for the rising tube to maintain its coherence. In
the corresponding figure in their paper one can even see the formation of a
vortex street in the wake, but the authors do not mention this
fact. \citet{Dorch:1999fk} carried out another 3D experiment, this time
simulating the rise of an $\Omega$-loop, and obtained that the apex of the
tube behaves as in the simple case of a 2D twisted tube.
\citet{abbett2000}, finally, carried
out 3D simulations of rising $\Omega$-loops also using the anelastic
approximation. Their results concur with those of the 2D simulations of
\citet{moreno1996} and \citet{emonet1998} for tubes with an amount of twist
at or above the critical limit obtained by those authors. For subcritical
twist, however, \citet{abbett2000} showed that the fragmentation of the $\Omega$-loop
depends on the loop's curvature; the higher the curvature, the smaller the
magnetic flux loss from the main body of the rising tube. All these early 3D simulations,
however, necessarily had a comparatively reduced number of grid points. Their
Reynolds number, therefore, was most likely not very high. 

Our aim in the present paper is to study the physics of the rise of a buoyant
magnetic flux tube through an adiabatically stratified medium in three
spatial dimensions with high spatial resolution of the tube and its boundary
layers, thus reaching comparatively high values of the Reynolds number. We
try to study in some detail the general features of the rise and also address
the question of the fragmentation of the tube and of the amount of magnetic flux
retained in its main body when 3D and comparatively high Reynolds number are used.
In order to diminish the numerical diffusivity, we performed the simulations 
using the FLASH code \citep{Fryxell:2000rr} with its AMR 
feature.  The governing system of 
equations, the numerical methods and the simulation setup are detailed in 
Section~\ref{sec:eq}. The global properties of the flux tube 
evolution are described in Section~\ref{sec:descrip}. In 
Section~\ref{sec:res}, we study the dependence of the tube fragmentation on the
curvature (Sections~\ref{sec:frag}) and on the amount of twist
(Sections~\ref{sec:twist}) for different levels of spatial resolution, or, more
precisely, of mesh refinement of the highly-structured subdomains. In
addition, Section~\ref{sec:struc} describes the differences in the structure
and dynamics of the $\Omega$-loop apex for the various simulations. In
Section~\ref{sec:lift}, we analyze the global effects of the aerodynamic lift force
on the legs of the $\Omega$-loops. Finally, Section~\ref{sec:conc} contains a
discussion and conclusions. 

\section{Setup of the simulations}\label{sec:eq}

We carry out a series of simulations for a simple compressible and electrically
conducting medium.  As equation of state we choose the ideal gas law ($p=\cal{R}
\rho T / \mu$), with $\rho$, $p$ and $T$ the mass density, gas pressure, and
temperature respectively, $\cal{R}$ the universal gas constant and $\mu$ the
atomic mass per particle. For a fully-ionized hydrogen plasma 
with realistic mass ratios, $\mu=0.5$. Since we are not concerned with the 
thermodynamics of the problem, the temperature scale is not important and we use $\mu=1$.  The system is governed
by the compressible ideal MHD equations. In conservation form and using MKS units, they are

\begin{eqnarray}
\frac{\partial \rho}{\partial t} &=& - \nabla \cdot(\rho \,{\bf v}) \label{eq:cont},\\
\frac{\partial (\rho\, {\bf v})}{\partial t} &=& - \nabla \cdot\left(\rho\,{\bf v}\otimes{\bf v} - \frac{{\bf B}\otimes{\bf B}}{\mu_0}\right) -  \nabla p_{T} +\rho\,{\bf g} \label{eq:moment}, \\
\frac{\partial (\rho\, e)}{\partial t} & = & - \nabla \cdot\left[{\bf v}(\rho\,
  e+p_{T})-\frac{{\bf B}({\bf v}\cdot{\bf B})}{\mu_0}\right] +  \rho \,{\bf g}\cdot{\bf v}\label{eq:ener}, \\
\frac{\partial{\bf B}}{\partial t}&=&\nabla \cdot ({\bf v} \otimes {\bf B}-{\bf B}\otimes {\bf v}), \label{eq:induction}
\end{eqnarray}

\noindent where $\otimes$ denotes the tensor product and ${\bf v}$, ${\bf B}$, ${\bf g}$ 
stand for the velocity field, magnetic field and gravitational acceleration, respectively.
Cartesian coordinates $(x,y,z)$ are adopted, with $z$ being height, i.e.,
${\bf g} = - g\; {\bf e_z}$ with $g>0$. 
$p_{T}$ is the total (magnetic + gas) pressure and $e$ is the specific total
energy, i.e., the sum of the kinetic and internal energies per unit mass and $B^2
/ (2\mu_0 \rho)$. 

\subsection{Numerical method}\label{sec:num}

The simulations were carried out using the FLASH 2.5 code \citep{Fryxell:2000rr}. 
This code implements a Riemann solver that is formally second-order accurate in 
time and space. The advective terms are discretized using a slope limited 
Total Variation Diminishing (TVD) scheme and the time-stepping is performed 
using an explicit Hancock-type scheme. 

Effects associated with thermal, viscous or ohmic transport phenomena have
not been explicitly taken into account in the MHD equations
\ref{eq:cont}-\ref{eq:induction}. Yet, artificial diffusive effects are
always present in all simulations as a result of 
the diffusion inherent in the numerical scheme. In our case
numerical diffusion is due to the TVD scheme acting to limit gradients at the
level of the numerical grid spacing. In astrophysical problems, the most
general situation is to have very large physical viscous and magnetic
Reynolds numbers, the solar interior being but one example thereof. To get as
close as possible to the astrophysical values in the numerical simulation,
one has to diminish the numerical diffusive effects, which requires
increasing the spatial resolution of the grid, thus yielding higher effective
Reynolds numbers. To that end, we use the AMR capability of FLASH to increase
the grid resolution in regions where it is required. Briefly, using AMR, the
cartesian domain is divided into adjacent square or cube blocks, each block
consisting, by default in the code, of $8\times 8$ (2D) or $8 \times 8 \times
8$ (3D) grid cells. All the blocks are checked to determine if they should be
redefined within a time interval which must be much smaller than the time
that the tube needs to cross one block, i.e., we set to check it every ten
timesteps. If the normalized second-order spatial derivative of the absolute
field strength exceeds some fixed threshold in any grid cell, the resolution
of the corresponding block is doubled using second-order interpolation. Then,
the original block is split into four (2D) or eight sub-blocks (3D),
increasing the refinement level by one. The reverse process happens when the
second order spatial derivative of the absolute field strength is smaller
than some threshold for all four (2D) or eight (3D) sub-blocks.

For further details on the FLASH code and its AMR algorithm, the 
reader is referred to \citet{Fryxell:2000rr} and the PARAMESH website  
\hfill\break
\hbox{http://www.physics.drexel.edu/\~{}olson/}
\hfill\break
\hbox{paramesh-doc/Users\_manual/amr.html}

\subsection{Initial and boundary conditions}\label{sec:cond}

The initial plane-parallel background is an adiabatically 
stratified polytrope described by the following density, 
pressure and temperature profiles:
\begin{equation}
\rho_e(z) = \rho_o\;\xi^{n-1}\quad ;\quad
p_e(z) = p_o\;\xi^n  \;;\label{eq:stratpres}
\end{equation}
\vspace{-2mm}
\begin{equation}
T_e(z) = T_o\;\xi\;, \label{eq:stratemp}
\end{equation}
where
\begin{equation}
\xi(z) = 1+\frac{z_o-z}{n\, H_{p_o}} \;;\label{eq:xi} 
\end{equation}

\noindent 
$\rho_o$, $p_o$ and $T_o$ are constant values, $z_o$ is a reference height, and 
$n$ is the polytropic index. The entropy profile for the stratification
given through (\ref{eq:stratpres}) - (\ref{eq:xi}) fulfills the following relation:
\begin{equation}
\frac{d s_e}{d \log p_e} = c_V\, \left(1 - \gamma \,\frac{n-1}{n}
\right)\;,\label{eq:entropy}
\end{equation}

\noindent with $\gamma$ the specific heat ratio (chosen here $\gamma = 5/3$)
and $c_V^{}$ the constant specific heat. To prevent the development of
convective cells in the experiment we choose an isentropic stratification,
i.e., we set \hbox{$n=\gamma/(\gamma-1) = 2.5$} so that the right-hand side of
Equation~\ref{eq:entropy} vanishes.  ${H_p}_o$ is the pressure
scale-height at $z=z_o$, as follows immediately from Equations~\ref{eq:stratpres} and
\ref{eq:xi}. 
We use $H_{p_o}$, $\rho_o$, $p_o$, and $T_o$ as units of length, density,
pressure, and temperature, respectively.  For the velocity, we choose as unit
the isothermal sound speed at $z_o$, namely, $(p_o/\rho_o)^{1/2}$.
As magnetic field unit we choose $B_o = (\mu_0 p_o)^{1/2}$. 
The non-dimensional units allow to apply these simulations in various problems 
where the stratification of the atmosphere is close enough to adiabatic. In the Sun, this 
problem could be implemented everywhere in the convection zone as long as it is not 
too close to the photosphere. By way of example, for the conversion of variables one 
could use the following set of values for the units, which are adequate to the deep 
convection zone: $H_{p_o} =  5.7\, 10^4$~km, $T_o= 2.2\, 10^6$~K, $p_o=6.5\, 10^{13}$~erg~cm$^{-3}$,
$\rho_o=0.22$~g~cm$^{-3}$.

Our aim is to have an initial distribution of magnetic field, pressure and
density that leads to the development of a rising $\Omega$-loop.  To that
end, we follow in the steps of \citet{Fan:2001kv}, and \citet{archontis2004},
and many later references that use similar distributions. The initial
magnetic flux tube is taken to be axisymmetric with the axis lying 
along the $y$-axis at a height $z=z_{o}$. In cylindrical coordinates
$(r,\phi,y)$ centered on the tube axis, the longitudinal (${\bf B_{l}}$) and
azimuthal (${\bf B_{t}}$) components of the magnetic field have the following
form:

\newcommand{\expmrsq}{{\exp\left(-\frac{r^2}{R^2}\right)}}
\newcommand{\exprsq} {{\exp\left( \frac{r^2}{R^2}\right)}}
\newcommand{\expmrsqy}{{\exp\left[-\frac{r^2}{R(y)^2}\right]}}
\newcommand{\exprsqy} {{\exp\left[ \frac{r^2}{R(y)^2}\right]}}

\begin{eqnarray}
{\bf B_{l}}&=&B_c \;\expmrsq  \;  {\bf
  e}_{y} \label{eq:blong}\;,\\ 
\noalign{\vspace{2mm}}
{\bf B_{t}}&=&B_c \; \expmrsq \; \lambda\,
\frac{r}{R}\;{\bf e}_{\phi} \;, \label{eq:btrans} 
\end{eqnarray}

\noindent where $R$ provides a measure for the radius of the flux tube and
$B_c$ is the initial magnetic strength along the tube axis. This prescription 
leads to twisted magnetic field lines that lie on the cylindrical surfaces 
$r={\rm const}$. For all simulations in this paper, $R=0.12$ and $B_c=0.2$, 
corresponding to a local plasma $\beta = 50$. 
\noindent The magnetic field is set to zero beyond $r=2R$. At this radius the
exponentials in Equations~\ref{eq:blong} and \ref{eq:btrans} are already below
$0.02$; the small resulting tangential discontinuity for the field is
dynamically unimportant.  
$\lambda$ in Equation~\ref{eq:btrans} is the dimensionless twist parameter. The
pitch of the twisted field lines is given by $(\Delta y)_{\rm pitch}  = 2\, \pi\,R /
\lambda$, independently of $r$.  As part of our parameter study, simulations
with different values of $\lambda$ were carried out.

The magnetic field distribution given by Equations~\ref{eq:blong} and
\ref{eq:btrans} leads to a Lorentz force pointing  inward the radial
direction. One can easily compensate for it by correcting for the gas
pressure, as follows:
\begin{eqnarray}
p(t=0) = p_e + p'\qquad \nonumber\\
\noalign{\noindent with\vspace{2mm}}
p' = - \frac{B_l^2}{2\,\mu_0} \left[1\;+\;\lambda^2\,\left(\frac{r^2}{R^2} \,-\,
       \frac{1}{2}\right)\right] \;, \label{eq:pressure_correction}
\end{eqnarray}
so, after applying this correction, there is still force equilibrium. Now, in the 3D
simulations, we are interested in studying the rise of 
$\Omega$-loops. To induce their development, we then introduce
a density perturbation localized around $y=y_0$ with spatial extent $\chi$:
\begin{equation}
\rho' = \rho_e  \left[ \left(\frac{p}{p_e}\right)^{1/\gamma} \hskip -2mm-
  1\right] \exp\left[-\frac{(y-y_0)^2}{\chi^2}\right]\;.
\label{eq:chi}
\end{equation}
and add it to the background stratification: 
\begin{eqnarray}
\rho(t=0) = \rho_e + \rho' \;,  \label{eq:density_perturbation}
\end{eqnarray}

\noindent maintaining the pressure as given in Equation~\ref{eq:pressure_correction}. 
Note that the entropy throughout the computational domain is kept
unchanged, which is different to \citet{Fan:2001kv} and
\citet{archontis2004}. Yet, the results published in this paper are not
dependent on the exact nature of this perturbation, whether isentropic or
isothermal. For a discussion of these perturbations see \citet{moreno1983}.
From Equation~\ref{eq:chi} we see that the tube is neutrally buoyant
($\rho' \to 0$) for $|y - y_0| \gg \chi$, so at time $t=0$ it is 
in complete force equilibrium there.  On the other hand, the perturbation
(\ref{eq:chi}) leads to a density deficit of order $|\rho'/\rho_e| \approx
(\gamma\beta)^{-1}$ around $y=y_0$, which is adequate to start the dynamical
evolution of the tube \citep[see][]{moreno1983}. From the Gaussian shape of
the density deficit (Equation~\ref{eq:chi}), we can estimate the radius of curvature
of the resulting $\Omega$-loop as follows: in this paper we are going to pursue the
loops until their apex has risen across a distance of a few pressure
scale-heights $H_{p_o}$. Hence, we expect the radius of curvature of the
rising loop to be of order 
\begin{equation}
R_c = \hbox{O}\,\left[{\chi^2 \over 2\,
    H_{p_o}}\right] \label{eq:radius_curvature} 
\end{equation}
at its apex. Consequently, we 
will use $\chi$ as the parameter controlling the radius of
curvature in the rising loops.  When $\chi \to \infty$, the initial condition
leads to an uniformly buoyant tube that rises maintaining a horizontal shape.

Concerning the simulation domain, the size of the 3D box was either $2.6
\times 10.4 \times 2.6$ or, for some experiments, $2.6 \times 15.7 \times
2.6$. The value $2.6$ in the transverse directions $x$ and $z$ provides 
enough room to accommodate a thin flux tube ($R=0.12$) inside it, even after
it expands along the rise. In the $y$ direction, we
need room to fully include the developing $\Omega$-loop, whose expected
radius of curvature is as given in Equation~\ref{eq:radius_curvature}. Hence the
elongated shape of the box along $y$ and also the larger $y$-size
for the 3D simulations with the highest $\chi$. In
the 2D experiments, the box was $2.6 \times 2.6$ in all cases. The center of
the initial tube was always located at $z=z_o = 11.1$, but the distance of
that point to the top and bottom lids of the box was varied slightly to
accommodate for the needs of each simulation (like the distance descended by
the lower segments of the tube). The experiments are finished when the top
layers of the tube reached $z \ge 12.3$. The pressure contrast between those two
levels ($z=11.1$ and $z=12.3$) is $5.2$, which corresponds to $1.6$ pressure
scale-heights.  The density contrast is $2.7$.

Table~\ref{tab:sim2} summarizes the different experiments carried out for this
paper. The main parameters we vary, following the explanations just given,
are $\chi$, i.e., the parameter associated with the radius of curvature (4th column in the table), the
twist parameter $\lambda$ (5th column), and the number of levels of
refinement allowed in the AMR algorithm over the basic, non-refined grid (6th
column). We label the experiments (first column) using those parameters. For example,
$\chi_{1.5}^{}\lambda_{1/8}^{}H$ is a simulation with $\chi=1.5$, $\lambda=1/8$ and ({\it
  H\,}) the {\it highest} number of refinement levels (in fact, AMR=2) over
the basic grid. Other experiments have a single level of refinement (AMR=1,
indicated with an ``M'') or no refinement (AMR=0, indicated with an
``L''). $\chi_\infty$ is the tag used for the case when $\chi \to \infty$ in
Equation~\ref{eq:chi}; this corresponds to uniform buoyancy along the
$y$-axis which leads to a horizontal rising tube, i.e, a 2D experiment.
Further columns in the table contain the box size (2nd column), the number of
grid points if the whole box were covered by a grid at the highest level of
AMR resolution in that experiment (3rd column) and, by way of results, the
amount of flux retained in the tube toward the end of the calculated
evolution (9th column) and the effective diameter of the head of the tube at
time $t\sim50$ (10th column). The table also contains a representative value
of the Reynolds number of each experiment (7th column) which we see varies as
a function of the initial parameters, in particular of the number of AMR
levels. Given that the viscosity in the model is numerical (the equations solved do 
not explicitly include a viscous term), one cannot directly compute the Reynolds 
number from characteristic lengths and flow speeds. We use the procedure described 
in Appendix~\ref{sec:rey} to estimate Reynolds number. From this Reynolds number one can estimate 
the viscosity following the expression $\nu = (L_c v_c)/Re$, where $L_c$ is the 
the characteristic length, i.e., the radius (see 10th column), and $v_c$ is the characteristic 
velocity. For the latter, one can use the rise speed of the emerging tube, which is of order 
0.3 in the various simulations.  

Concerning boundary conditions, we choose periodicity for all variables on 
all side boundaries.  As for the top and bottom lids, we use open boundary
conditions there, i.e., the plasma can go into and out of the box; further,
the density is extrapolated across those boundaries and the gas pressure is
deduced considering local hydrostatic equilibrium.

\begin{deluxetable*}{lccccccccc}
\tablecaption{\label{tab:sim2} List of the simulations described in the paper.
  The columns contain: Name of the simulation; size 
  of the box; equivalent number of grid points at the highest resolution
  level; radius-of-curvature parameter $\chi$; 
  twist parameter $\lambda$; AMR levels; estimated Reynolds number
  ($Re$) and dimensionless viscosity ($\nu$). In addition the following results are listed for 
  $t=50$: percentage of 
  flux retained inside the tube ($\Fret$), and effective diameter of the head 
  of the tube.}  
\tablehead{ \colhead{ name } & \colhead{box size} &
  \colhead{grid points} &\colhead{$\chi$} & \colhead{$\lambda$} & \colhead{AMR} &
  \colhead{Re} & \colhead{$\nu$} & \colhead{$\Fret$(\%)} & \colhead{D$_{\rm head}$} } \startdata
$\chi_\infty^{}\lambda_0^{}H$ & 2.6, 2.6 & $512,512$ & $\infty$ &0 & 2 & - & - & 8
& 0.20 \\
$\chi_\infty^{}\lambda_{1/16}^{}H$ & 2.6, 2.6 & $512,512$ & $\infty$ & 1/16
& 2 & 190 & $3\, 10^{-4}$ & 20 & 0.40 \\ 
$\chi_\infty^{}\lambda_{1/8}^{}L$& 2.6, 2.6 & $128,128$ & $\infty$ & 1/8 & 0
& 3.5 & $10^{-2}$ & 5 & 0.36 \\
$\chi_\infty^{}\lambda_{1/8}^{}H$ & 2.6, 2.6 & $512,512$ &  $\infty$ & 1/8 & 2 &  190 & $3\, 10^{-4}$ & 47 & 0.68\\ 
$\chi_\infty^{}\lambda_{1/4}^{}H$ & 2.6, 2.6 & $512,512$ & $\infty$ & 1/4 & 2 & 190 & $4\, 10^{-4}$ & 72 & 0.74\\  
$\chi_2^{}\lambda_{1/16}^{}H$ & 2.6, 15.7, 2.6 & $512,3072,512$ & $2.1$ 
&   1/16 & 2 & 290 & $10^{-4}$ & 32 & 0.42 \\ 
$\chi_2^{}\lambda_{1/8}^{}L$ & 2.6, 15.7, 2.6 & $128,768,128$ & $2.1$ & 1/8 & 0 & 7.5 & $6\, 10^{-3}$ &  20 & 0.46\\  
$\chi_2^{}\lambda_{1/8}^{}M$ & 2.6, 15.7, 2.6 & $256 , 1536 , 256$ &  $2.1$ & 1/8 & 1 &  100 & $10^{-3}$ &  41 & 0.60 \\
$\chi_2^{}\lambda_{1/8}^{}H$ & 2.6, 15.7, 2.6 & $512 , 3072 , 512$ & $2.1$ &  1/8 & 2 & 290 &  $2\, 10^{-4}$ & 48 & 0.54 \\ 
$\chi_2^{}\lambda_{1/4}^{}H$ & 2.6, 15.7, 2.6 & $512 , 3072 , 512$ & $2.1$ &   1/4 & 2 & 290 & $3\, 10^{-4}$ & 72 & 0.74\\ 
$\chi_{1.5}^{}\lambda_0^{}L$ & 2.6, 10.4, 2.6 & $128, 512 , 128$ &  $1.5$
&0 & 0 & - & - & 16 & 0.26\\  
$\chi_{1.5}^{}\lambda_0^{}H$ & 2.6, 10.4, 2.6 & $512 , 2048 , 512$ & $1.5$ & 0 & 2 & - & - & 18 & 0.18\\  
$\chi_{1.5}^{}\lambda_{1/16}^{}H$ & 2.6, 10.4, 2.6 & $512 , 2048 , 512$  & $1.5$ & 1/16 & 2 &  300 & 
$2\, 10^{-4}$ &  36 & 0.48 \\ 
$\chi_{1.5}^{}\lambda_{1/8}^{}L$ & 2.6, 10.4, 2.6 & $128 , 512 , 128$ &$1.5$ &   1/8 & 0 & 53 & $10^{-3}$ & 50 & 0.72 \\  
$\chi_{1.5}^{}\lambda_{1/8}^{}M$ & 2.6, 10.4, 2.6 & $256 , 1024 , 256$ & $1.5$ & 1/8 & 1 & 100& $6\, 10^{-4}$ & 51& 0.62\\ 
$\chi_{1.5}^{}\lambda_{1/8}^{}H$ & 2.6, 10.4, 2.6 & $512 , 2048 , 512$ & $1.5$ & 1/8 & 2 & 300 & $2\, 10^{-4}$ & 49 & 0.66 \\ 
$\chi_{1.5}^{}\lambda_{1/4}^{}H$ & 2.6, 10.4, 2.6 & $512 , 2048 , 512$ & $1.5$ & 1/4 & 2 &  300 & $2\, 10^{-4}$ & 72 & 0.68\\ 
$\chi_{1}^{}\lambda_{1/16}^{}H$ & 2.6, 10.4, 2.6 & $512,2048,512$  &  $1$ & 1/16 & 2 & 340 & $10^{-4}$ & 38 & 0.50 \\
$\chi_{1}^{}\lambda_{1/8}^{}L$ & 2.6, 10.4, 2.6 & $128 , 512 , 128$  & $1$ & 1/8 & 0 &  75 & $9\, 10^{-4}$ & 60 & 0.70\\  
$\chi_1^{}\lambda_{1/8}^{}M$ & 2.6, 10.4, 2.6 & $256 , 1024 , 256$  &  $1$ &  1/8 & 1 &100 & $6\, 10^{-4}$ & 58 & 0.62\\ 
$\chi_1^{}\lambda_{1/8}^{}H$ & 2.6, 10.4, 2.6 & $512 , 2048 , 512$ & $1$ & 1/8 & 2 &  340 & $2\, 10^{-4}$ & 55 & 0.60 \\
$\chi_1^{}\lambda_{1/4}^{}H$ & 2.6, 10.4, 2.6 & $512 , 2048 , 512$ & $1$ & 1/4 & 2 &  340 & $2\, 10^{-4}$ & 74 & 0.68 \\
\enddata
\end{deluxetable*}

\section{General evolution of the 3D tube structures}\label{sec:descrip}

\begin{figure*}
\begin{center}
\includegraphics[width=1.0\textwidth]{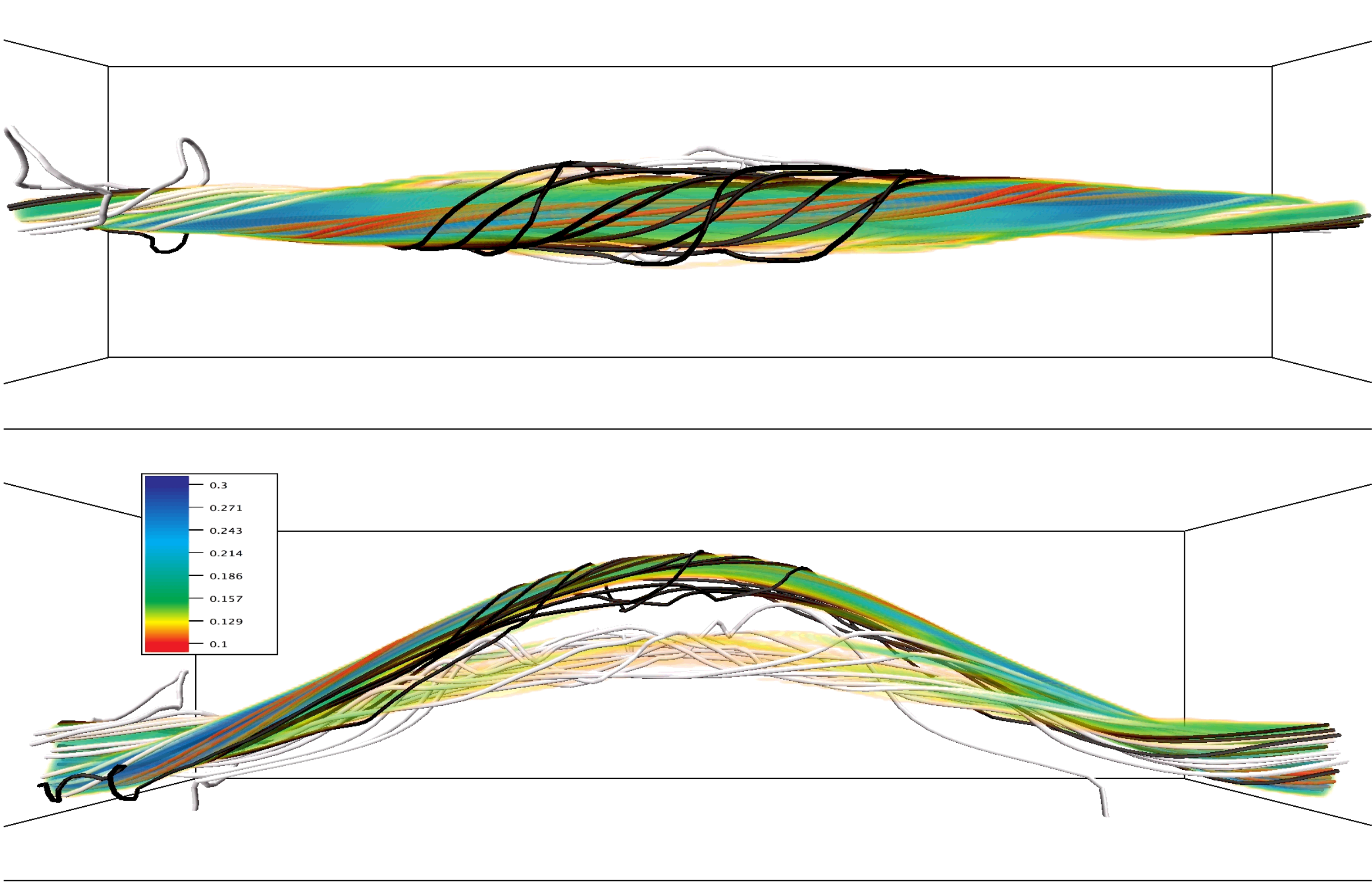}
\end{center}
\caption{\label{fig:3db} {\footnotesize Illustration of the $\Omega$-loop 
structure. The magnetic field strength is shown in three dimensions with a blue-yellow 
color scale for simulation $\chi_{1.5}^{}\lambda_{1/8}^{}H$ at  $t=62$ 
including a top view (upper panel) and a side view (lower panel). Different
sets of magnetic field lines are also shown, with colors explained in the text.
The accompanying online figure allows 3D navigation: it has two 
iso-contours, a yellow semitransparent one at $|B| = 0.2$, for better visualization of 
the tail of the tube, and one at $|B|=0.3$ to explore the head region and the same 
set of field lines as the figure. }}
\end{figure*}

We would first of all like to show a number of general features of the 3D
evolution of a rising flux tube when using the highest level of Adaptive Mesh
Refinement.
To that end we take experiment $\chi_{1.5}^{}\lambda_{1/8}^{}H$, which has an
intermediate level of curvature of the resulting $\Omega$-loop ($\chi=1.5$)
and of the field line twist ($\lambda=1/8$), and in which a relatively high
Reynolds number, $Re \approx 300$, is reached. Figure~\ref{fig:3db} 
shows a general 3D view of the system at an intermediate stage of the rise 
as seen from the top and side boundaries (top and bottom panels,
respectively). In the figure, a volume rendering of the magnetic field
strength is presented and a number of representative field lines added. A few
remarkable features are worth mentioning. In the bottom panel, a seemingly
double structure of the tube is apparent: there are two separate arched
strands, a green one at the top and a yellow one at the bottom.
The actual $\Omega$-loop that contains most of the axial magnetic
flux is just the green strand at the top; the yellow lower strand is part of 
the complicated wake of the tube described below. The roots, in turn, have 
a twisted appearance, with a clear antisymmetry between the right and left 
footpoints (see top panel of the figure) that is explored in more detail 
in Section~\ref{sec:lift}. The expansion following the rise across successive
pressure scale-heights is also apparent: the magnetized region
at the top is wider than the roots. The field lines drawn in the figure,
finally, correspond to a few of the general patterns encountered in the tube:
smooth, twisted field lines go through the main body of the tube structure
(red lines, visible directly on top of the volume rendering, specially in the
upper panel) and through the magnetized region in 
the wake (white lines). Some of the field lines drawn go 
through the boundary layer at the periphery of the tube (black lines): the 
field vector along the stretch of these lines going through the top is 
essentially transverse, which fits with the fact that the highest level of 
twist in a rising and expanding tube is always reached toward its top. 

\begin{figure*}
\begin{center}
\includegraphics[width=0.95\textwidth]{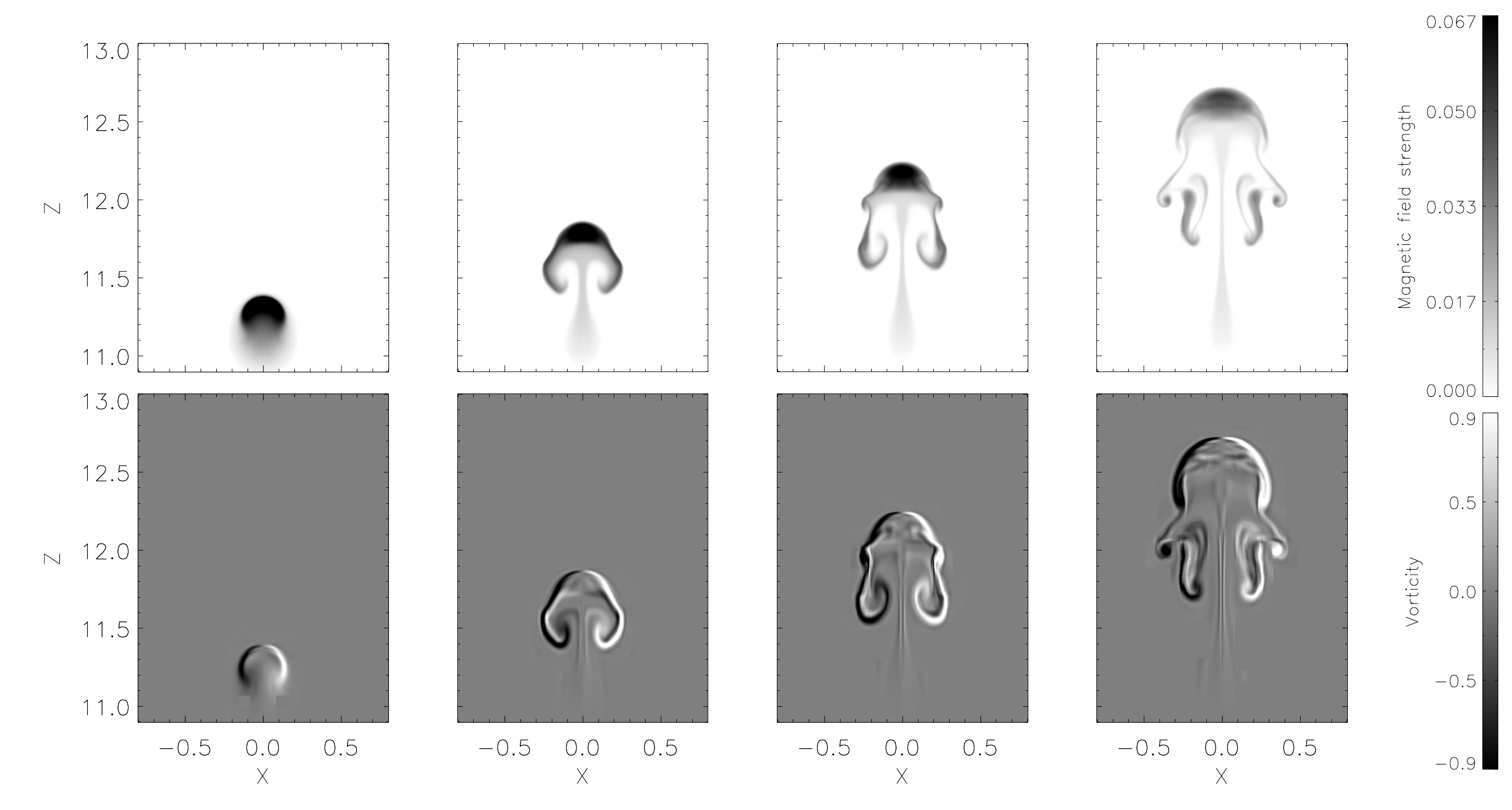}
\end{center}
\caption{\label{fig:vort} 
 {\footnotesize Longitudinal component of the 
 magnetic field (top) and of the vorticity (bottom) are shown in 
 grey-scale in a $xz$ section at the apex of the $\Omega$-loop 
 for simulation $\chi_{1.5}^{}\lambda_{1/8}^{}H$ at times $13$, $35$, $48$ 
 and $62$, from left to right respectively.}}
\end{figure*}

The 3D view provided in Figure~\ref{fig:3db} can be completed through the 
time evolution of the cross section of the same tube presented in 
Figure~\ref{fig:vort} for times $t=13$, $35$, $48$ and $62$. In the figure,
magnetic field strength (top row) and vorticity (bottom row) maps are drawn on a
vertical cut that coincides with the midplane of the box perpendicular to the
initial axis of the tube. We confirm that most of the magnetic flux is
concentrated at the very top of the structure. The general appearance and
evolution of the structures shown are strongly reminiscent of those described
by \citet{moreno1996}, \citet{emonet1998} and \citet{mark2006} using
2D simulations. We recall a few salient features obtained in those papers and
reproduced here: in the initial stages (two leftmost panels), the rising
portion of the flux tube has a {\it mushroom} shape, akin to an air bubble
rising in water \citep[e.g.,][]{daviestaylor50, collins65, parlange69,
  wegenerparlange73, hnatbuck76, ryskinleal84b, christovvolkov85}. Within the
ascending portion of the tube, vorticity is generated at the boundary layer
between the tube and the ambient flow, in fact with opposite sign on the
either side of the head (bottom row in the figure). Magnetic flux is dragged
along the sides of the tube into a trailing pair of counter-rotating
vortices.  Later in time the wake is fragmented into smaller and smaller
vortices because the Reynolds number increases in time due to the expansion
of the tube \citep{mark2006}. The AMR device allows to see
this fragmentation in much more detail than could be obtained in the original
work of \citet{emonet1998}. Inside the head of the tube, on the other hand,
the plasma and magnetic field are executing a twisting oscillation, with
small velocity compared to the global rise speed.  This description of the
time evolution of the flux tube structure generally applies to other
simulations with comparable levels of refinement even when the twist and
curvature vary. The cases with untwisted flux tubes, instead, follow a
somewhat different pattern and their evolution is described in the following.

\begin{figure*}
\centering
\includegraphics[width=0.95\textwidth]{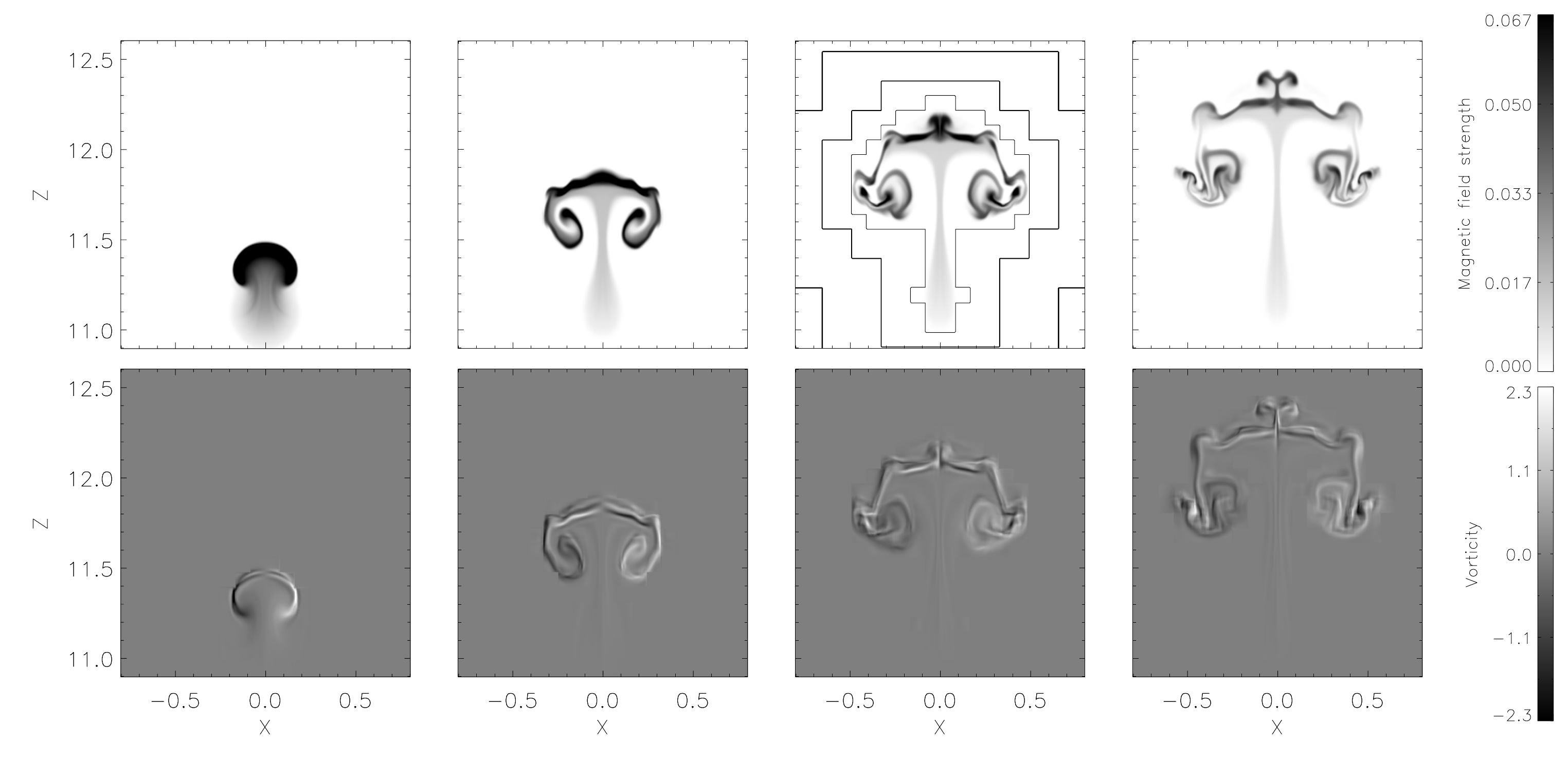}
\caption{\label{fig:evol} {\footnotesize Longitudinal component of the
    magnetic field (top row) and of the vorticity (bottom row) are shown in
    grey-scale in a $xz$ section at the apex of the $\Omega$-loop for
    simulation $\chi_{1.5}^{}\lambda_0^{}H$ at times $18.2$, $39.6$, $54.5$
    and $64.6$, from left to right respectively. The black contours shown in
    the top-third panel are the different AMR levels, i.e., where the tube
    is, the AMR level is the highest, and the outer AMR level is the
    lowest. }}
\end{figure*}

An example of a simulation for an initially untwisted $\Omega$-loop with a
relatively high Reynolds number is experiment $\chi_{1.5}^{}\lambda_0^{}H$,
shown in Figure~\ref{fig:evol}. The appearance of the tube in the early phase
(leftmost panel) resembles the 2D simulations by~\cite{Shussler1979} and
\citet{longcope1996}. After rising a distance of roughly one to two initial
tube diameters, the tube has split into two counter-rotating vortex rolls. As
the magnetic structure continues its rise, the top of the structure is
pressed by the ambient flow into a horizontal bar shape. Thereafter, the bar
is disrupted by an instability producing small vortices, possibly a
Rayleigh-Taylor instability as described by \citet{tsinganos1980}. This
process spawns a flux {\it tubelet} emanating from the top of the tube (3rd
and 4th column of the figure). As the tubelet rises above the main portion of
the tube, it suffers a similar fate as the original tube, in that it splits
into two counter-rotating rolls. In both cases, when a pair of
counter-rotating rolls has been created, their sideways separation from each
other is due to the action of the aerodynamic lift \citep{longcope1996}.
This whole phenomenon is probably also the reason for the structures shown in
the 3D simulation of \citet{Dorch:1998db} (see their Figure~3, middle row),
which are reminiscent of what we obtain here.

\begin{figure}
\centering
\includegraphics[width=0.49\textwidth]{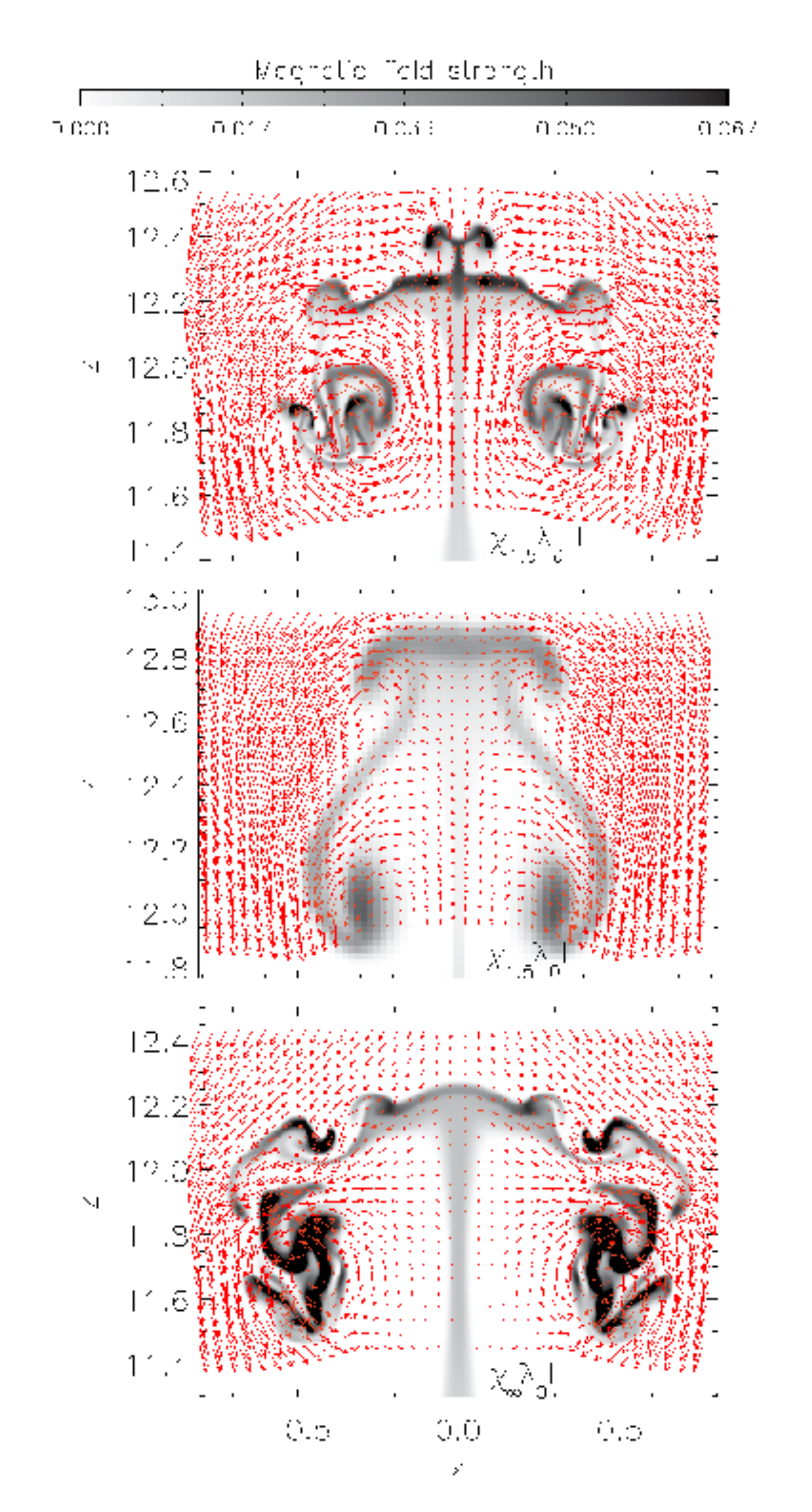}
\caption{\label{fig:bzv2} {\footnotesize Cross-sectional structure of the
    tube at $y = 0$ for the three simulations without twist as follows:
    $\chi_{1.5}^{}\lambda_0^{}H$ at $t=64.6$ (top panel),
    $\chi_\infty^{}\lambda_0^{}H$ at $t=47$ (middle panel), and
    $\chi_{1.5}^{}\lambda_0^{}L$ at $t=72.7$ (bottom panel). The longitudinal
    component of the magnetic field is shown in grey-scale. The red arrows
    correspond to the projection onto the vertical plane of the relative
    velocity field with respect to the apex.}}
\end{figure}

The possible influence of the initial parameters on the bar structure is
briefly explored in Figure~\ref{fig:bzv2}, which compares maps of the
longitudinal component of the magnetic field in the vertical midplane of the
box for the three experiments carried out for tubes with no initial twist
($\lambda=0$): apart from the $\Omega$-loop experiment just described
($\chi_{1.5}^{}\lambda_0^{}H$, top panel), we present an experiment for a
straight horizontal rising tube ($\chi_\infty^{}\lambda_0^{}H$, middle panel)
and an experiment like the one at the top but with no AMR
($\chi_{1.5}^{}\lambda_0^{}L$, bottom panel). In all three experiments, the
head of the tube expands, becoming horizontally elongated like a bar. This
bar shape is possibly associated both with the larger amount of work needed 
to expand vertically (due to gravity) and with the lower pressure of the 
external flow on the sides due to the elementary Bernoulli effect 
\citep[e.g.,][]{parker1979}.  
The sideways expansion and deformation is not counteracted by magnetic 
tension due to the lack of twist in the tube. Within the bar, the motion is
negligible except for the slow horizontal expansion of the tube in a similar
way as happens in 2D simulations with twist, which have a negligible relative
motion inside the head \citep{emonet1998}.  The corresponding
velocity shear between the tube and the outside plasma in the bar yields a
vorticity distribution with opposite signs above and below it (as visible in
the lower row of Figure~\ref{fig:evol}).  In the two cases with higher Reynolds number (top
and bottom panels of Figure~\ref{fig:bzv2}), the bar is later deformed with
subsidiary wiggles. In fact the bar shows strong shear and has a density
deficit in its interior, so the wiggles are possibly produced by a
combination of the Rayleigh-Taylor and Kelvin-Helmholtz instabilities. 

\section{The retention of magnetic flux as a function of curvature, Reynolds
  number and field line twist}\label{sec:res}

The central aim of this paper is to study which parameters influence the
amount of flux lost by the rising $\Omega$-loops using our 3D,
high-resolution framework. From the possible parameters we select two which
have been shown in previous studies to be particularly important in the
process, namely, the field line twist \cite[studied for rising horizontal
tubes using 2D simulations by][]{moreno1996, emonet1998}, and the
curvature of the $\Omega$-loop \cite[studied using low-resolution 3D
simulations by][]{abbett2000}. We then check if that dependence is
substantially modified when increasing the spatial resolution \citep[as 
done in 2D by][]{mark2006}.

There are various ways to measure the flux loss and the tube
fragmentation. One method was developed by \citet{moreno1996} and
\citet{emonet1998}, who measure the amount of flux inside the main
body of the tube (i.e., excluding the rolls in the wake and the tail) and
compare it with the flux of the initial tube. \citet{longcope1996}
and \citet{abbett2000}, on the other hand, measure the 
sideways expansion of the vortices resulting from the splitting of the
initial tube. We follow the first method and calculate the
magnetic flux retained within the head of the tube at the apex of the
$\Omega$-loop ($\Fret$).  To that end, we choose representative snapshots
for each experiment and integrate the value of $B_y$ in the region inside the
head of the tube; for this measurement, we define the latter as the region
where $B > \Beq$  where $\Beq$ is the equipartition value with the
kinetic energy: 

\begin{equation}\label{eq:equipartition}
B_{eq}^2/(2 \mu_0) = \rho\, v^2/2 \;. 
\end{equation}

\noindent For simulations with low AMR, we use a less strict condition, 
namely, $B > 0.25 \Beq$ since 
the evolution of the tube leads to large relative kinetic energy with respect 
to the apex and  in some cases, as shown later, the main body of the 
head of the tube remains together (see Section~\ref{sec:struc} for details).  
That value is subtracted from the total longitudinal flux in the
tube at the initial time and the result is used as an approximation for the
flux loss from the tube for the given run.

\subsection{Dependence of flux retention on tube curvature}
\label{sec:frag}

A particularly interesting dependence to explore is with the curvature
of the axis of the rising tube. \citet{abbett2000} reported that, in the 3D case of rising
$\Omega$-loops, the fragmentation of the rising tube was smaller the
larger the curvature of the loop axis. Given the early date of all
those results, the spatial resolution available to them was quite
limited, specially the 3D ones. Our approach with Adaptive Mesh
Refinement allows for much higher spatial resolution (hence a much
larger Reynolds number). In the following we would like to test the
flux loss dependence on curvature for different AMR regimes. 

\begin{figure*}
\centering
\hbox to \hsize{
\includegraphics[width=0.49\textwidth]{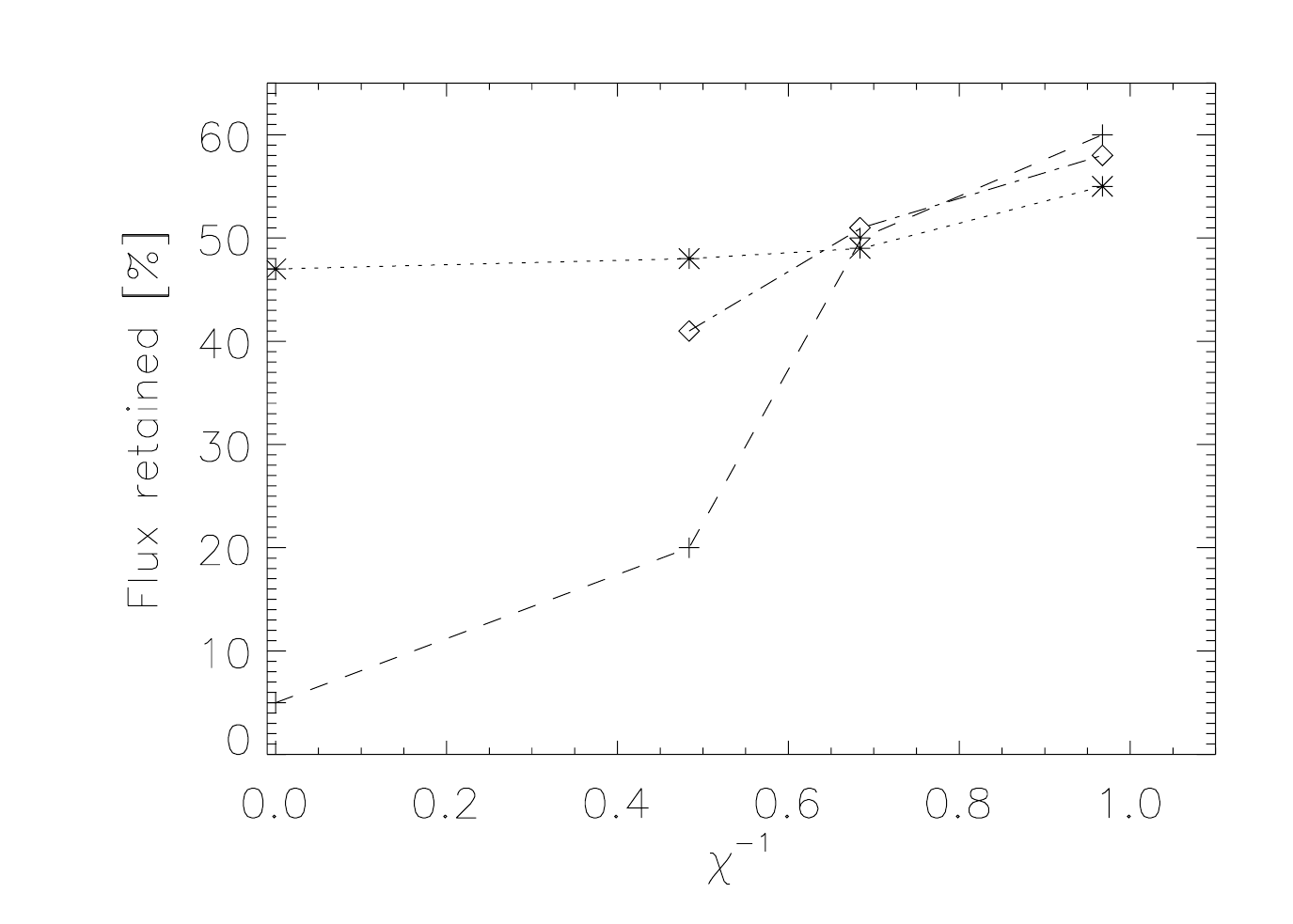}
\includegraphics[width=0.49\textwidth]{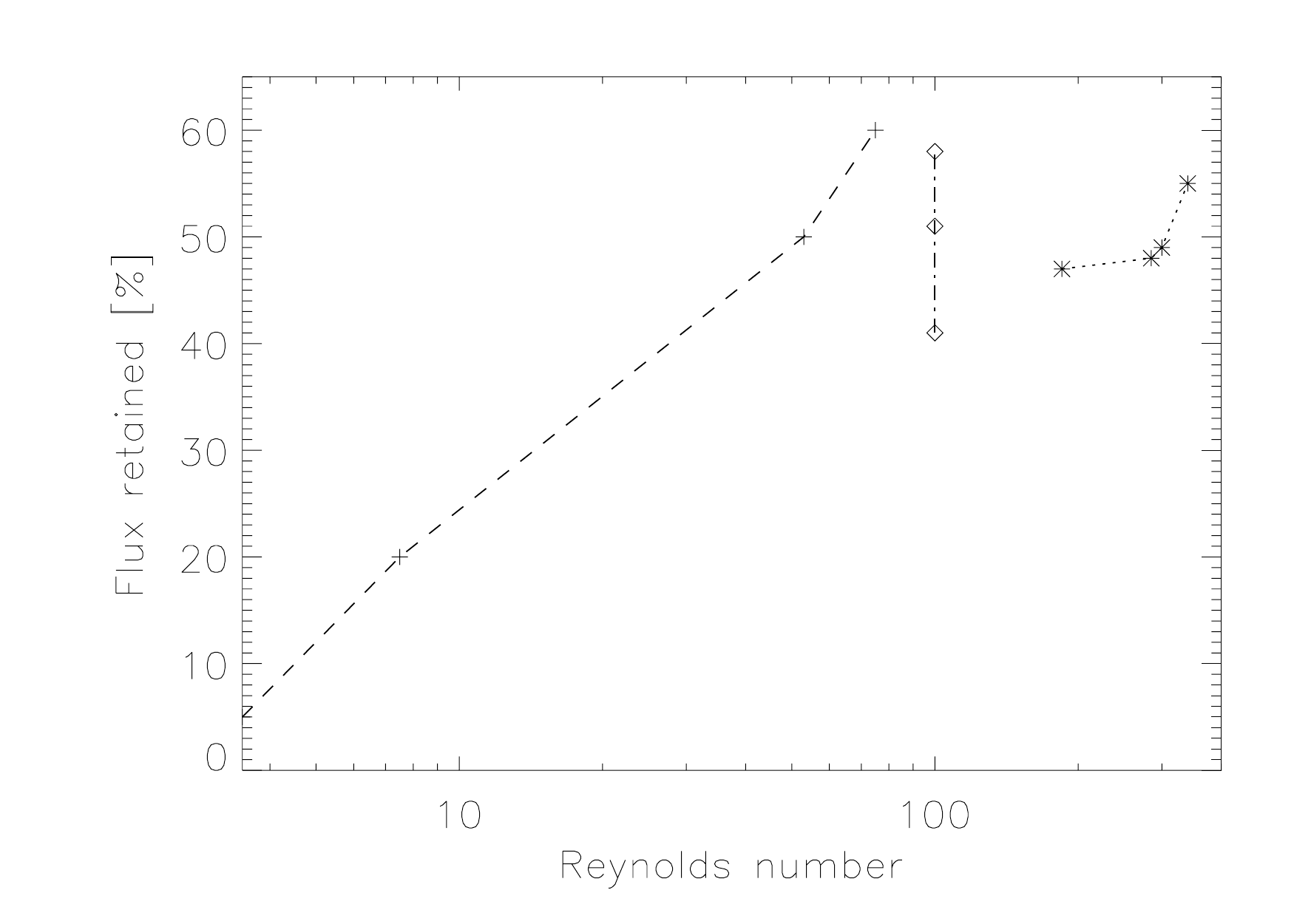}}
\caption{ {\bf Left panel:} $\Fret$ plotted
  as a function of the curvature parameter ($\chi^{-1}$) for simulations with
  AMR=0 ($+$ symbols, dashed line), AMR=1 ($\diamond$, dot-dashed line) and
  AMR=2 ($\ast$, dotted line). The cases with $\chi^{-1} = 0$ correspond
  to 2D simulations. All cases considered in this figure have magnetic twist
  $\lambda=1/8$. 
{\bf Right panel:} Magnetic flux retained within the head of the tube ($\Fret$) as
a function of Reynolds number for the same experiments as on the left
panel. 
Also here, the lines link cases with the same level
of AMR (namely, from left to right: AMR $= 0, 1, 2$). 
\label{fig:flucur}}
\end{figure*}

Figure~\ref{fig:flucur} (left panel) displays a plot of the flux retained in
the head of the tube, $\Fret$, as a function of the curvature parameter
$\chi$ for simulations with a common value of the field line twist, namely,
$\lambda=1/8$. We have linked with lines the experiments with the same 
AMR level. For the lowest AMR level (dashed line), we find 
that flux retention increases with curvature, as reported by \citet{abbett2000}.  
In fact, comparing the cases
with the extreme values of curvature ($\chimone = 0$, i.e., horizontal tube,
left end of the curve and $\chimone = 0.95$, right end of the curve) we see 
a difference of close to $60\%$ in the amount of flux retained in the
head of the tube. Yet, going to cases with higher levels of AMR we see that the 
effect of curvature  diminishes: the dash-dotted line (for AMR=1) 
shows a much less marked dependence, while the cases with the highest AMR 
(dotted line) show very little dependence with the curvature parameter. The amount
of flux retained in those cases is in the range $40\% - 60\%$, which, as we 
will see in the next subsection, can be taken as representative for the value of 
twist used in all simulations in this figure. 

The same trends as described above for the left panel of
Figure~\ref{fig:flucur} are revealed in the right panel which shows the
effective Reynolds number against the amount of flux
retained (the line style indicates the same values of the AMR used in the
left panel). By comparing the two panels we see that, in the low AMR regime, the effective Reynolds
number is highly dependent on the curvature. For the high AMR cases, instead,
the effective Reynolds number, as well as the flux retained, are only weakly
dependent on the curvature. The behavior and properties of the tube for low
and high diffusion regimes are clearly different.  For AMR=0, the Reynolds number
increases with curvature because of the variation in the size of the tube
head. Instead, when using a higher number of AMR levels, the Reynolds
number is seen to increase basically through the decrease in the width of
the boundary layer.

In summary, we are able to reproduce the dependence of flux retention on 
curvature as reported by \citet{abbett2000} only when the system is 
sufficiently diffusive. When higher AMR is used and sufficiently high-Re is 
achieved, the dependence of these properties on curvature is substantially diminished.
We tentatively conclude that the curvature does not play any major role when the 
diffusion is low enough. 

\begin{figure}[h]
\centering
\includegraphics[width=0.49\textwidth,height=0.59\textwidth]{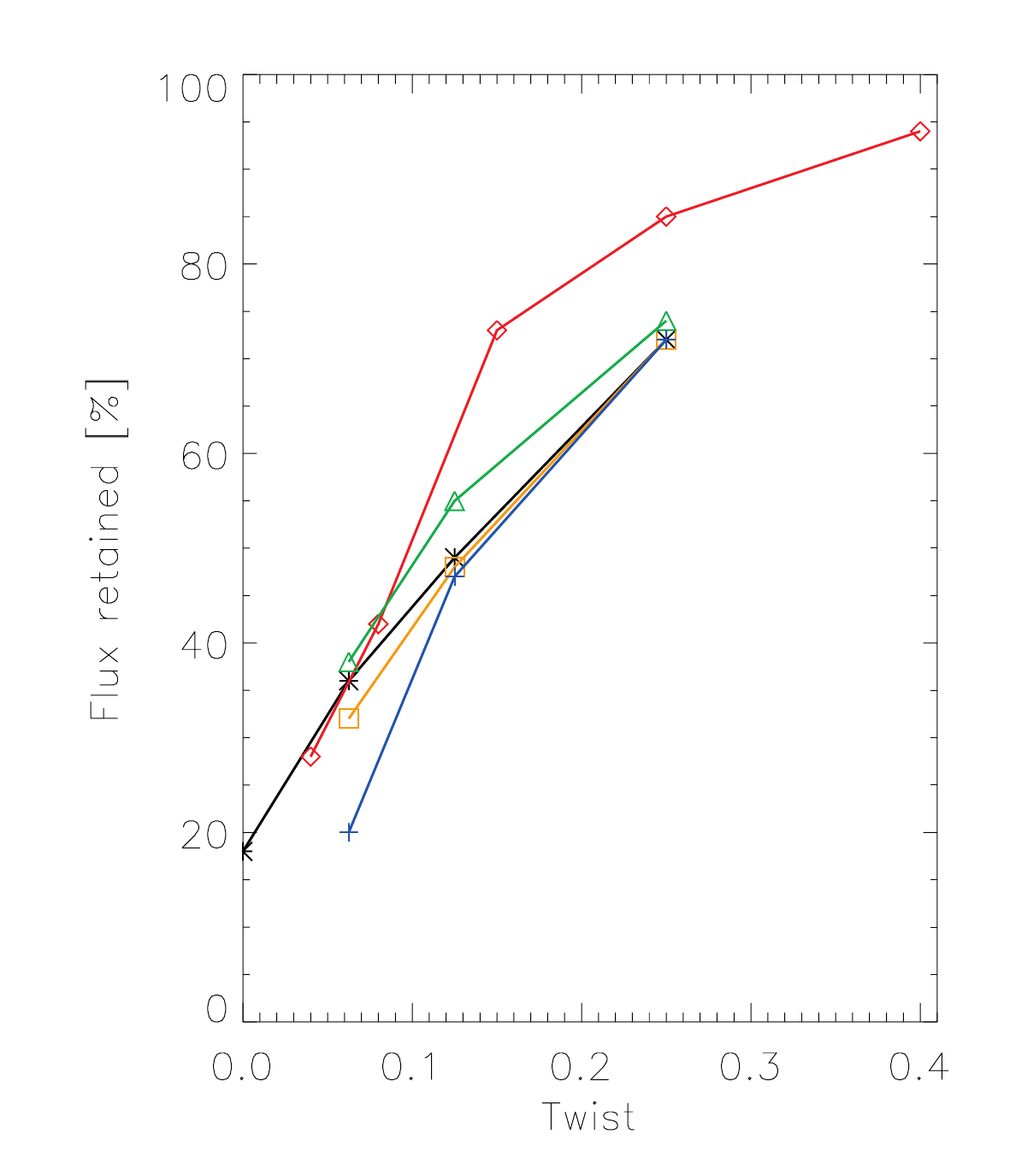}
\caption{\label{fig:flutw} {\footnotesize $\Fret$ as a function of
    field line twist $\lambda$. The cases shown are: 2D experiments,
    i.e., $\chimone=0$ ($+$ symbols, blue line); experiments with 
    $\chimone=0.48$ ({\small $\Box$}, orange line); 
    $\chimone=2/3$ ($\ast$, black line);  
    $\chimone=1$ ({\small$\triangle$}, green line). All the
    foregoing experiments cases were calculated with 
    the highest AMR level. For comparison, 2D results from \citet{mark2006}
    are also plotted ($\diamond$, red line).}}
\end{figure}

\subsection{Dependence of flux retention on twist}
\label{sec:twist}

Here we would like to focus on the twist itself. Figure~\ref{fig:flutw} shows
$\Fret$ as a function of the field line 
twist for $\lambda$ between $0$ and $1/4$. On the basis of the results of the
previous section,  
we concentrate here on the cases with the highest AMR level
(hence, with the highest Reynolds numbers). The lines link cases with the
same value of the curvature parameter, namely, from the case with highest
curvature ($\chimone=1$, triangles, green line) through $\chimone=2/3$ 
(asterisks, black line) and $\chimone=0.48$ (squares, orange line),
down to the cases with horizontal tubes ($\chimone=0$, crosses, blue
line). For comparison, we are also overplotting the corresponding 2D
results of \citet{mark2006} (their Figure~5), shown here with diamonds linked
by a red line: those values were calculated with a much higher level of
AMR (hence, of the Reynolds number) as was possible in the 3D calculations of
the present paper. We also note that none of our 3D calculations in the
figure are started with strongly twisted tubes: taking the highest twist used
in the figure, $\lambda = 1/4$, and the pitch value given in
Section~\ref{sec:cond}, we see that the field lines only complete some $3.5$
turns in the whole length of the box along the $y$-direction for the
experiments with box size $10.4$.

A clear result from Figure~\ref{fig:flutw} is that, for each given fixed
curvature, $\Fret$ dramatically increases with increasing magnetic field line
twist following a pattern of the general shape obtained by
\citet{moreno1996} (their Figure~3) and \citet{mark2006} irrespective
of the curvature of the $\Omega$-loop.  We note that the results of the 3D
experiments for each $\lambda$ cluster around a single value: the major
deviation from this behavior appears for the smallest value of the twist
($\lambda = 1/16$), but even there the percentage of flux retained
is less than a factor two difference between the different cases. This
reinforces the conclusion reached in the last section that above a
spatial resolution threshold the curvature of the $\Omega$-loop is not the
primary factor determining the amount of magnetic flux retained by the rising
tube. The possible variation when going to higher resolution levels can be
gauged by comparing the 2D results shown in the figure (crosses, red line) 
with those of \citet{mark2006}: for $\lambda = 1/4$, for
instance, the high-Re results are a factor roughly $1.2$ above the
moderate-Re ones.  

\subsection{Multi-parametric study of the tube structures} \label{sec:struc}

We study in the following the variation of the cross-sectional structure of
the rising magnetized domain when changing the main parameters in the study while
keeping the others fixed. We carry out three comparisons: dependence with the
twist parameter for an $\Omega$-loop with $\chi=1.5$ and the highest AMR
level (Section~\ref{sec:structure_variation_with_twist}); dependence with
curvature for the twist parameter fixed at $\lambda=1/8$ for calculations
with the highest AMR (Section~\ref{sec:highre}); and, finally, the same 
dependence but for calculations with no AMR (Section~\ref{sec:lowre}).

\begin{figure*}
\centering
\includegraphics[width=0.75\textwidth]{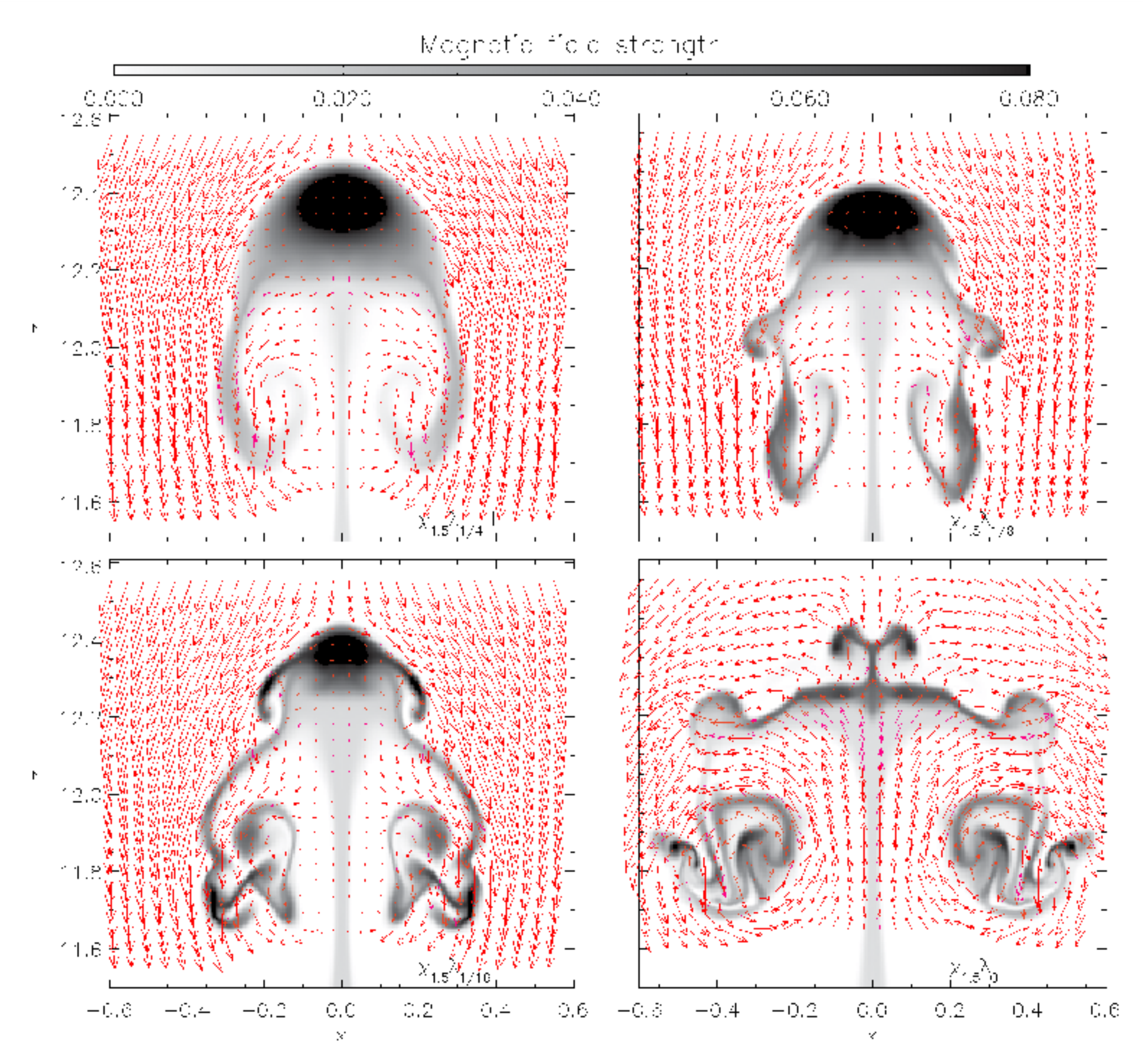}
\caption{\label{fig:bz3b} {\footnotesize Dependence of the
    cross-sectional structure of the tube on the level of twist
    ($\lambda$). The layout is the same as in
    Figure~\ref{fig:bzv2}. 
The simulations from left to right and top to bottom are
$\chi_{1.5}^{}\lambda_{1/4}^{}H$ at $t=59 .2$, 
$\chi_{1.5}^{}\lambda_{1/8}^{}H$ at $t=54.5$, 
$\chi_{1.5}^{}\lambda_{1/16}^{}H$ at $t=59.2$, 
and $\chi_{1.5}^{}\lambda_0^{}H$ at $t=64.6$.}}
\end{figure*}

\subsubsection{Twist}\label{sec:structure_variation_with_twist}

Figure~\ref{fig:bz3b} shows the distribution of the longitudinal magnetic
field $B_y$ in the $xz$ plane at the apex of the rising tubes for various
degrees of twist ($\lambda$) but fixing the curvature at $\chi=1.5$ and
always using the highest refinement level (AMR = 2). As one would expect from
Section~\ref{sec:twist}, one sees a reduction in the size of the {\it head}
of the tube for decreasing twist. Beyond that, however, our high-resolution
calculations allow to discern the high level of substructure one obtains when
going to less twisted cases. For those cases in which a head is maintained
(i.e., the three cases with $\lambda >0$), we see that the shedding of
vortices to the wake takes place earlier in the lower-$\lambda$ cases, with
at least a few vortex-shedding instances having taken place in the 
$\lambda = 1/8$ and $\lambda = 1/4$ experiments at the apex. 
Given the absence of net global vorticity in the cross section shown in the figure, the shedding 
of vortices is symmetric and no classical Von Karman vortex street is 
formed behind the tube \citep{emonet2001}, i.e., no instability leading 
to an alternation in the vortex shedding from either side develops for the 
duration of the current experiments.  On the other hand, one sees that
the elaborate structure shown in the bottom right panel here and studied in
Section~\ref{sec:descrip} (Figure~\ref{fig:bzv2}) only develops for untwisted
tubes: even the limited level of the transverse field of the case shown in
the bottom-left panel is sufficient to suppress the development of the 
vortex tubelets issued forward from the central bar.

\subsubsection{Curvature - high AMR}\label{sec:highre}

\begin{figure*}
\centering
\includegraphics[width=0.75\textwidth]{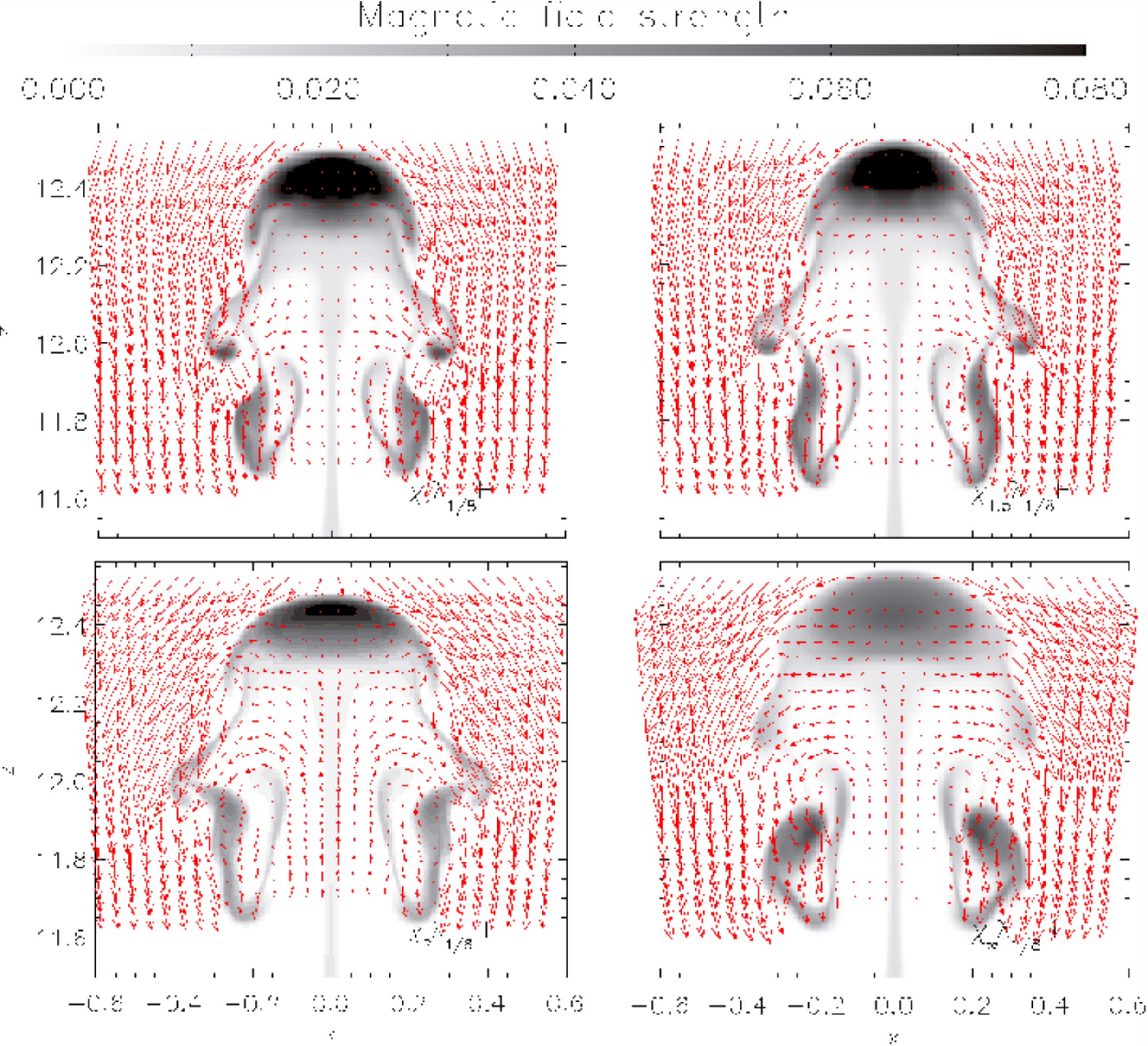}
\caption{\label{fig:bz3a} {\footnotesize Dependence of the
    cross-sectional structure of the tube on the curvature
    parameter ($\chi$). The layout is the same as in
    Figure~\ref{fig:bzv2}.  The simulations from left to right
    and top to bottom are 
$\chi_1^{}\lambda_{1/8}^{}H$ at $t=62$, 
$\chi_{1.5}^{}\lambda_{1/8}^{}H$ at $t=56.6$, 
$\chi_2^{}\lambda_{1/8}^{}H$ at $t=54.5$, 
and $\chi_\infty^{}\lambda_{1/8}^{}H$ at $t=49.8$. Compared to the dependence with twist, the 
tube structures are less sensitive to tube curvature.}}
\end{figure*}

In contrast to the strong changes seen when varying $\lambda$ in
Figure~\ref{fig:bz3b}, the cross-sectional structure of cases with different
curvature but constant $\lambda$ and high AMR is remarkably similar
(Figure~\ref{fig:bz3a}). This property is in agreement with the fact that the
amount of magnetic flux inside the tube ($\Fret$) varies only  little
with curvature when the spatial resolution is high enough 
(Section~\ref{sec:frag}). A small difference between the various
simulations shown in Figure~\ref{fig:bz3a}, however, is that the horizontal
expansion of the head of the tube is smaller the higher the curvature of the
$\Omega$-loop (i.e, the smaller the value of the $\chi$ parameter). This seems to
be a consequence of the fact that the relative rise of the crest of the
$\Omega$-loops compared with their flanks entails a stretching of the tube
along the axis which is larger the greater the curvature. This leads 
to an enhancement of the pressure deficit there which may
counteract the tube expansion as it rises, even if to a limited extent.

\subsubsection{Curvature - low AMR}\label{sec:lowre}

\begin{figure*}
\centering
\includegraphics[width=0.75\textwidth]{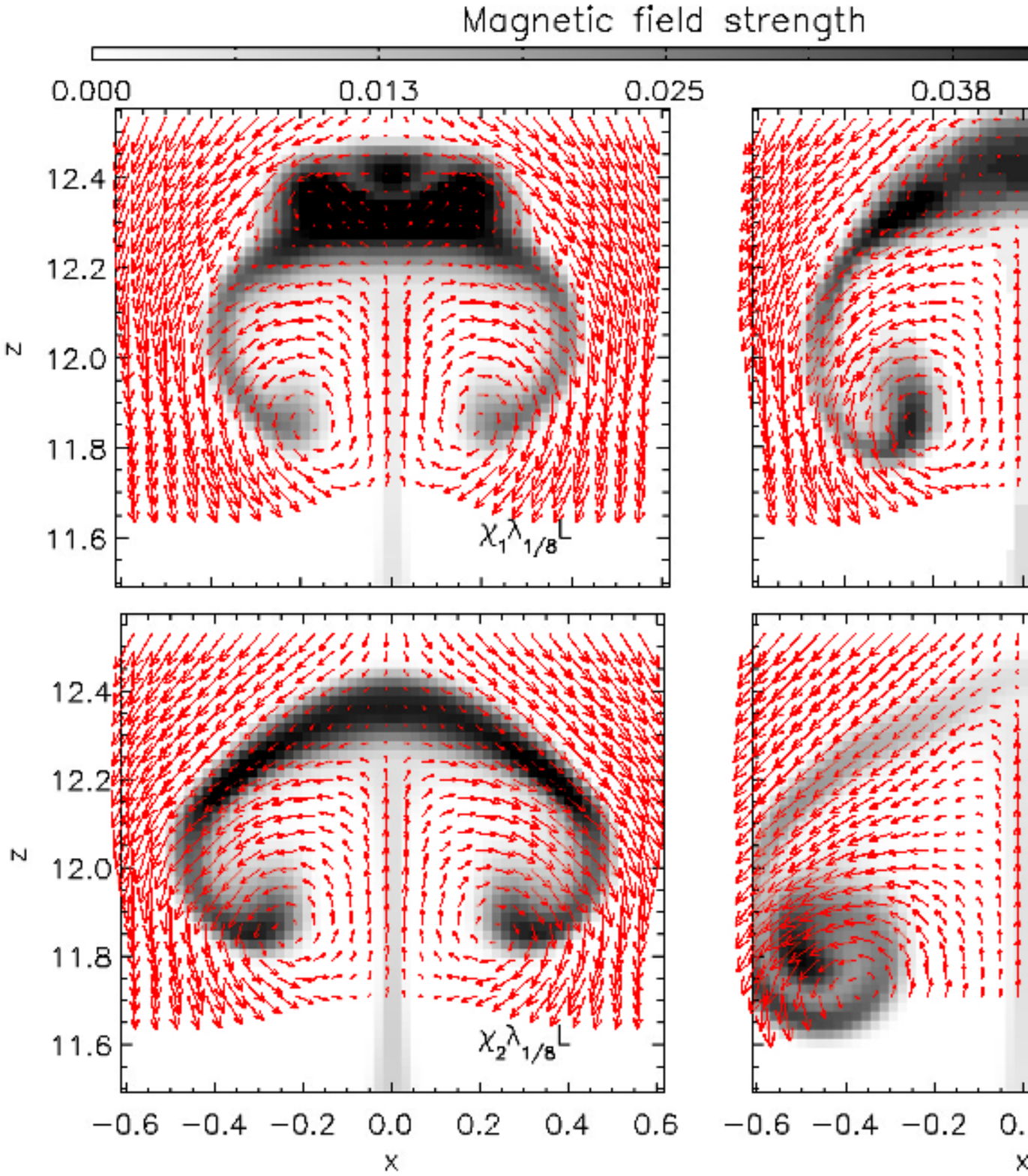}
\caption{\label{fig:bz3a_low} {\footnotesize Dependence of the
    cross-sectional structure of the tube on the curvature ($\chi$) for
    simulations without adaptive mesh refinement (AMR=0). The layout is the same as in
    Figure~\ref{fig:bzv2}. The simulations from left to right and top to
    bottom are $\chi_1^{}\lambda_{1/8}^{}L$ at $t=74$, 
    $\chi_{1.5}^{}\lambda_{1/8}^{}L$ at $t=73$, 
    $\chi_2^{}\lambda_{1/8}^{}L$ at $t=55$, 
    $\chi_\infty^{}\lambda_{1/8}^{}L$ at $t=73$. In contrast to the
    relatively high $Re$ regime, the structure of tube near the loop apex is
    very sensitive to the tube curvature.}}
\end{figure*}

The structures seen in the previous two subsections fit well with the results
on the magnetic flux retention by the head of the tube for the high-AMR
experiments seen in the previous section (Section~\ref{sec:twist} and
Figures~\ref{fig:flucur}   
and \ref{fig:flutw}). Looking at the dashed line in both panels of Figure~\ref{fig:flucur},
however, one wonders what may be causing the sensitive dependence on 
the curvature in the calculations without AMR. To check for this, we are
showing in this section the cross sectional distributions of the longitudinal
magnetic field for runs with different values of the curvature parameter
$\chi$ when switching off the Adaptive Mesh Refinement (AMR$=0$, 
Figure~\ref{fig:bz3a_low}). For all experiments, we keep $\lambda=1/8$, as
also used for Figure~\ref{fig:flucur}.

Figure~\ref{fig:bz3a_low} shows very different magnetic field 
distributions depending on the curvature. In this low--AMR regime,
magnetic diffusion is sufficiently dominant to significantly debilitate
the transverse field and hence the level of  magnetic twist, and 
the ability of the tube to remain coherent. In the 2D case there 
is a clear separation  of the tube into two vortex rolls which drift 
apart away from the central mid plane. In the 3D cases with high 
diffusion, instead, the two counterrotating vortices remain near 
the apex, where they are visible as two roundish dark features. 
The lift force would make those vortices move away from the 
central mid plane. However, in the figure we see that they stay 
nearer the apex the larger the curvature of the rising 
$\Omega$-loop. This may be due to the fact that 
the stretching of the matter along the axis of the loop that accompanies the rise
is more intense 
the larger the loop curvature. This leads to a relative gas pressure deficit
at the apex which counteracts the motion of the vortices away from the
midplane and which is more effective for more strongly curved loops. This
helps the head of the tube maintain some identity during the simulated time
interval in the low AMR cases with high axis curvature. In the cases with high 
spatial resolution of the previous section (Section~\ref{sec:highre}, 
Figure~\ref{fig:bz3a}), instead, the tension of the twisted transverse field inhibits 
any relevant dynamics inside the tube.

\section{Horizontal motion of the legs of the magnetic loop due to aerodynamic lift}
\label{sec:lift}

\begin{figure}[h]
\centering
\includegraphics[width=0.47\textwidth]{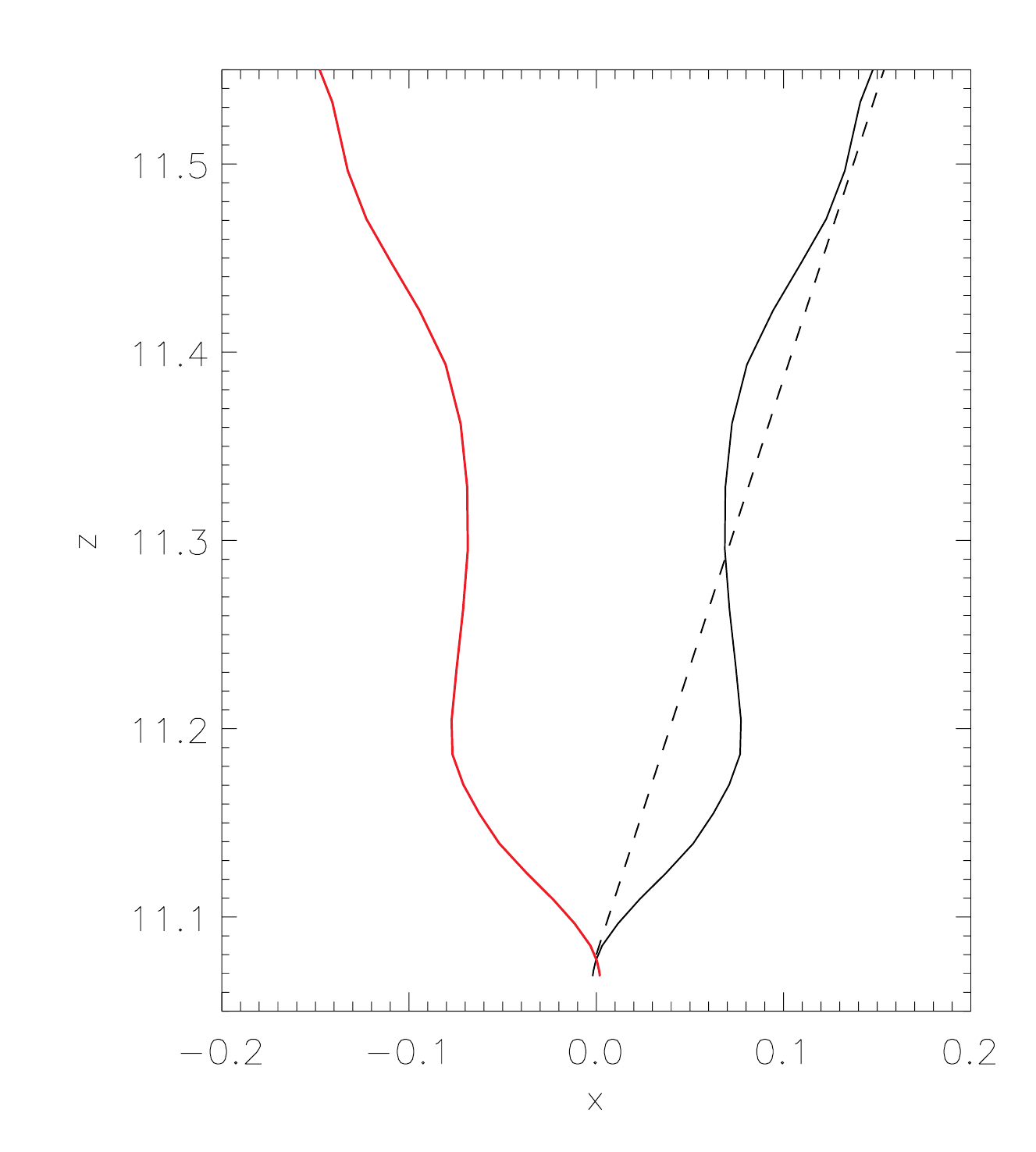}
\caption{\label{fig:lift1} {\footnotesize Black and red solid lines: 
trajectory of the center of the tube  
in the vertical $xz$ plane for simulation $\chi_{1.5}^{}\lambda_{1/8}^{}H$ at 
two equidistant positions with respect to the apex (red line at $y = - 2.7$ and black 
line at $y = 2.7$) Dashed line: simple trajectory calculated with fixed values of vorticity and
buoyancy as determined in the tube averaged within $t= [13,62]$ (see text for details).}} 
\end{figure}

In the simplified 1D image of a rising $\Omega$-loop as a thin, untwisted
magnetic tube, sometimes the motion of the loop can be assumed to be
contained in a vertical plane. To be sure, in the absence of effects
associated with solar rotation and when the tube does not have any net
vorticity anywhere along its length, all forces acting on the magnetized
plasma point either in the tangential or normal direction to the tube axis,
i.e., they cannot bring the tube out of a vertical plane if at any time it is
completely contained in one. In the present paper, however, there may 
appear vortex forces associated with any net vorticity arising in any 
stretch of the tube, and they may carry the tube out of the initial vertical plane.
In this section we show that this effect indeed takes place, in fact in an 
antisymmetric fashion for the two legs of the rising loop.

\subsection{Preliminary considerations}\label{sec:vorticity_preliminary}

\begin{figure*}
\centering
 \includegraphics[width=0.95\textwidth]{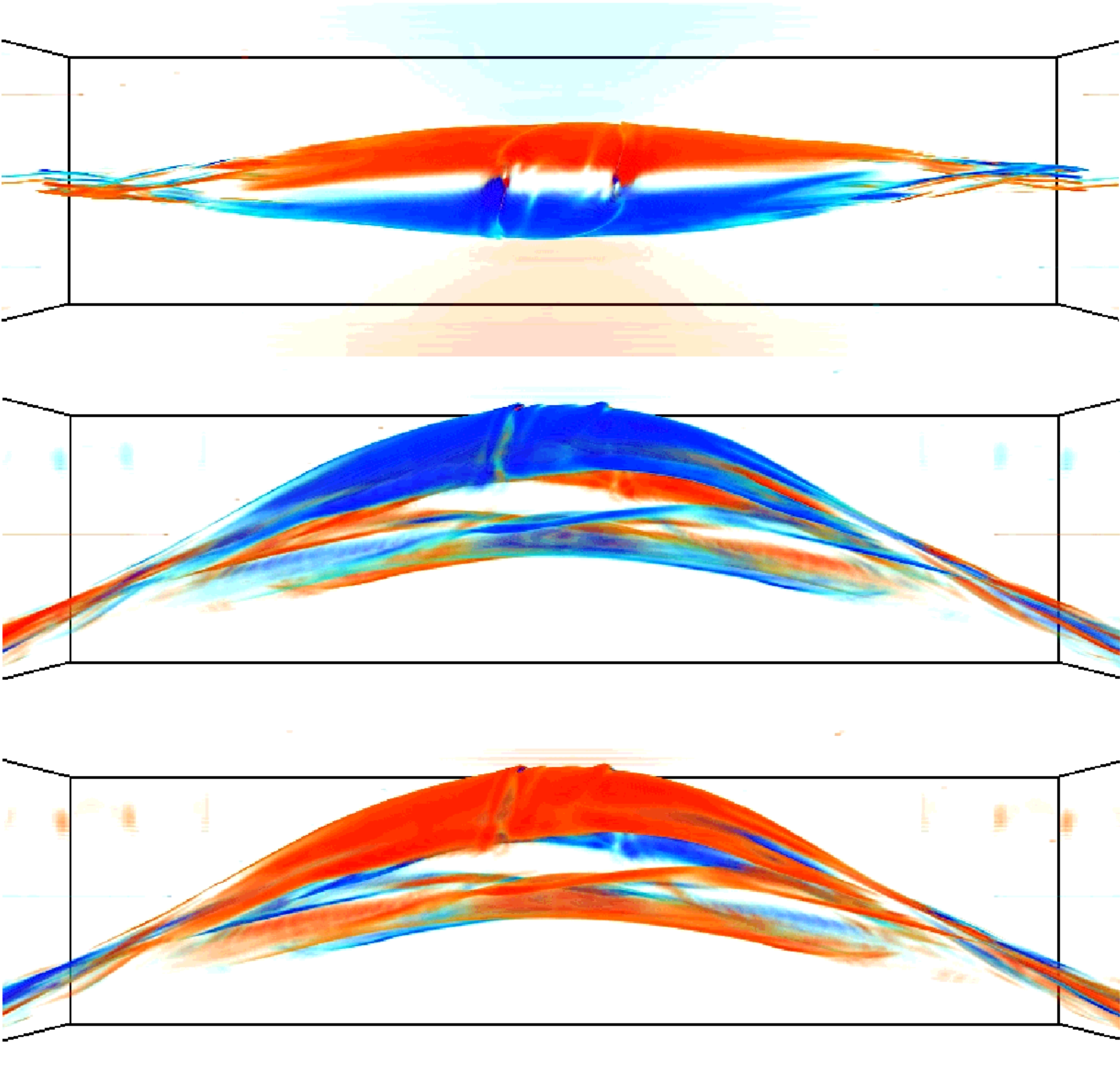}
\caption{\label{fig:vrtz3d} {\footnotesize 3D images of the $y$
    component of the vorticity for simulation
    $\chi_{1.5}^{}\lambda_{1/8}^{}H$ at $t=65.95$ 
    (color-scale) are shown for a $x$-view (top panel) and the
    opposite $x$-view (middle panel) and $z$-view (bottom panel), where
    red is negative and blue is positive $[-3.607, 3.607]$. The wake
    has an antisymmetric behavior between the left and the right
    wings.}}
\end{figure*}

To obtain a first impression, we recall that the axis of the initial tube is
contained in the vertical $yz$ plane at $x=0$; the sideways motion of the
axis away from that plane can be studied by pursuing the intersection of the
tube axis with the transverse vertical $xz$ plane across either leg of the 
rising loop midway between the apex and the roots. Figure~\ref{fig:lift1} 
shows the evolution of the intersection for the planes at $y=\pm 2.7$. 
In order to follow the center of the tube, we first locate it at
the apex of the $\Omega$-loop by searching for the minimum transverse
magnetic field. Second, we trace the field line going through that point
downward along the wings of the tube until they reach the chosen 
$y={\rm const}$ 
plane. The figure corresponds to experiment $\chi_{1.5}^{}\lambda_{1/8}^{}H$, 
in which, as we know from previous sections (e.g., Figure~\ref{fig:bz3b}, 
top-left panel), the tube retains most of the flux in the rising body instead of 
shedding it to the wake. The legs (solid lines in Figure~\ref{fig:lift1}) 
actually move off the initial $yz$ vertical plane in an antisymmetric way, 
following an inclined path with a superimposed oscillation sideways from 
it. The antisymmetry is also discernible in the 3D images displayed in 
Figures~\ref{fig:3db} and~\ref{fig:vrtz3d}. The latter figure shows the $y$ 
component of the vorticity for simulation $\chi_{1.5}^{}\lambda_{1/8}^{}H$ 
using three different views.  The upper panel shows a top view: we
clearly see the flanks of the tube being displaced in opposite directions on
either side of the crest. The two side views (middle and bottom panels) show
antisymmetries also in the vertical position of the main body of the tube and
in the wake. 

The basic reason for the right-left antisymmetry is that the legs of the loop
carry net vorticity pointing along the tube axis and of opposite sense on
either side of the apex.  The force associated with a vortex line of
strength $\mathbf{\Gamma}$ along its axis that moves with relative speed
$\mathbf{v}$ against the background leads to a lift acceleration given by 
$\mathbf{v}\times \mathbf{\Gamma}$ (e.g.,
\citealt{1932hydr.book.....L, 1995vody.book.....S}; see also
\citealt{longcope1996}). 
Later on in this section we will be discussing why the legs
of the tube develop net vorticity, in fact with vorticity vector pointing in
opposite senses on either side of the tube head. Given that fact, and noting
that the relative flow of the ambient medium seen by the tube points
downward, one obtains a vortex force pointing in the $x$ direction and, also
here, in opposite senses for each leg of the rising tube.

The motion of the axis off the central $yz$ vertical plane is thus a result
of the lift force in combination with the buoyancy and drag forces.  That
sort of motion is particularly simple in the case of a rectilinear horizontal
tube moving under the action of constant buoyancy, lift force and aerodynamic
drag \citep[]{emonet2001} for small values of the dimensionless parameter
\begin{equation}\label{eq:param_emonet_etal_2001}
\Psi = \pi \, R \, \Gamma^2 / (C_D G)\;,
\end{equation}
with $C_D$ the aerodynamic drag
coefficient (taken $C_D = 2$ below) and $G$, and $R$ the average
buoyancy acceleration, and the radius of the tube
cross section, respectively: in that case, the motion of the tube has uniform
velocity and the trajectory of its center is a straight line inclined to the
vertical by an angle $\theta$ given by $\Psi = \sin^2\theta/\cos\theta$. 
Our tube is not horizontal and does not move uniformly; yet, for the sake of
the comparison, we have included in Figure~\ref{fig:lift1} (dashed line) such a
trajectory for $\Psi=0.1$ (correspondingly, $\theta=18$ deg), which is the
average of $\Psi$ calculated for the tube elements represented in the
trajectory of the figure.

We can discern two main reasons for the appearance of net vorticity in the
rising tube:

\begin{enumerate}[(i)]
\item \label{enum:i} The action of the Lorentz force
  components associated with the
  varying amount of field line twist and cross sectional area one
  encounters when moving from the crest of the tube to the
  flanks. 

\item \label{enum:ii} The appearance of a trailing Von
  Karman vortex street behind the rising tubes: each time the tube sheds a
  vortex roll from its wake into the vortex street, the rest of the tube
  remains loaded with an equal amount of vorticity of opposite
  sign \citep{von1963aerodynamics,emonet2001}. 
\end{enumerate}

Those two effects are discussed in turn in the following subsections:

\subsection{Effect of the inhomogeneous distribution of cross-sectional
  radius and field line twist along the tube}\label{sec:inhomogeneity}

Broadly speaking, twisted tubes tend to equalize twist along their length
attempting to reach dynamical equilibrium: non-uniformities of the field line
pitch along the axis promote plasma spin-up around the axis, e.g., in the
form of a torsional Alfv\'en wave, that counteracts this lack of uniformity
\citep[see, e.g.,][]{longcope_klapper_1997,Longcope:2000yq,Fan:2009rt}. One
can illustrate this process by studying the azimuthal component of the
Lorentz force in simplified situations. \citet{longcope_klapper_1997} studied
the equations that govern the evolution of a thin flux tube with twisted
field lines and spinning motion around the axis. The thinness of the tube is
meant in the sense that only the lowest-order terms need be retained in a
radial expansion around the tube axis. Calling $A$ the cross sectional area
of the tube, they obtained an equation for the evolution of the angular
momentum around the axis as follows:

\begin{equation}\label{eq:thintube_lk}
{\partial\, (\omega \, A) \over \partial t} = A\, \gamma_2\,
{v}_A^2\, {\partial \,q \over \partial \, l}\; ,
\end{equation}

\noindent where $\partial / \partial l$ is the derivative along the axis of
the tube, $\omega$ the {\it spin} of the tube around its axis, $q$ the
magnetic field torsion defined as $q=({\bf J}\cdot {\bf B})/B^2$, ${v}_A$ an
average Alfv\'en speed in the tube cross section and $\gamma_2$ a form factor
related to the 2nd-order moment of the field distribution across the tube's
cross section (see \citealt{longcope_klapper_1997}, Equations~33 and A5). 
From Equation~\ref{eq:thintube_lk}, if the azimuthal component of the
magnetic field varies along the axis in such a way as to make $\partial
q/\partial l \ne 0$, then a spin-up of the tube follows that tends to
counteract this lack of uniformity.

Another comparatively simple situation is that of an axisymmetric flux tube with
straight axis, now without imposing any  condition on the size of
the tube radius. In this case, one can easily split the azimuthal component
of the Lorentz force into two parts: 

\begin{enumerate}[(a)]
\item \label{enum:a} the variation of the azimuthal
  field component as one slides in the axial direction leads to a radial
  component of the electric current. The latter, when combined with the axial
  field component, causes a force in the 
  azimuthal direction, i.e., an axial torque. This leads to spin-up (i.e. to
  change of the axial angular momentum per unit mass) that tries to make the
  twist uniform.

\item \label{enum:b} for tubes with
non-uniform cross section, the radial field component combined 
with the axial electric current that generally flows in twisted flux tubes
leads to another component of the force in the azimuthal direction, i.e.,
again, to an axial torque and to change of the axial angular momentum. 
\end{enumerate}

We further note that the axial angular momentum per unit mass at an arbitrary
radius of the simple axisymmetric tube is equal to the integrated amount of
vorticity in the circular cross section of that radius. 

In our simulations we can use neither one of the simplified avenues just
discussed: the magnetic tube is not thin; 
concerning the second simplified approach, the axis of our tube is not
straight and the initial axisymmetry is not strictly conserved along the
evolution. However, examining the results of the simulation we can clearly
conclude that the expansion experienced by the top part of the tube leads to
a local minimum of the azimuthal field component and of the pitch angle at
the apex inside of the main body of the tube (as apparent in 
Figure~\ref{fig:3db}, red field lines ). Also, as expected
from ideal MHD considerations, we see that the tube maintains along time the
negative helicity configuration of the initial condition. Those facts,
combined with the simplified considerations given in (\ref{enum:a}) above,
allow us to expect an azimuthal component of the Lorentz force with positive
(negative) sign in the left (right) wing of the tube. The corresponding
torque leads to creation of axial vorticity in the tube pointing toward the
apex, as measured in the experiment. Similarly, concerning (\ref{enum:b}) in
the list of the previous paragraph, the cross section of the tube is maximum
at its apex. One can check that the decrease of the cross section when moving
down the flanks of the tube combined with the tube's negative helicity also
leads to an axial torque that creates axial vorticity pointing toward the apex.
We think that it is the vorticity created through these mechanisms, combined
with the downward-pointing relative velocity of the surrounding medium with
respect to the tube (as explained in Section~\ref{sec:vorticity_preliminary}), that
results in a vortex force that causes the global sideways displacement in the
$x$-direction shown in Figure~\ref{fig:lift1}.

\begin{figure}[h]
\centering

\includegraphics[width=0.45\textwidth]{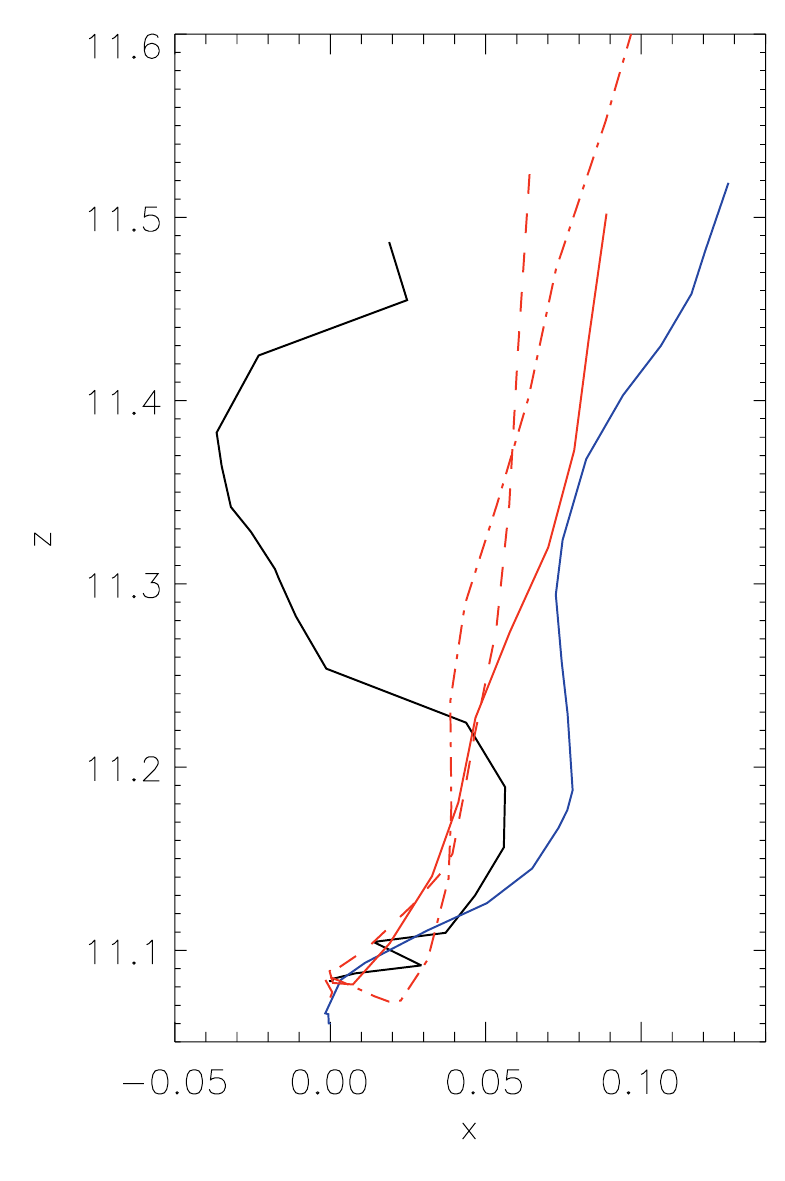}
\caption{\label{fig:lift2} {\footnotesize  Trajectory of the center of
    the tube for simulations with twist 1/4 (red), twist 1/8 (blue) and
    twist 1/16 (black). The solid lines are the trajectories in the vertical
    plane $y=3.1$ for cases with $\chi = 1.5$; 
the dash-dotted line is the trajectory in the plane $y=2.6$ for
    the case $\chi = 1$; finally, the dashed line
    is the trajectory in the plane $y=3.9$ for the case $\chi = 2.1$.}}
\end{figure}

\subsection{Formation of a Von Karman vortex street}\label{sec:vonkarman} 

\begin{figure*}
\includegraphics[width=0.95\textwidth]{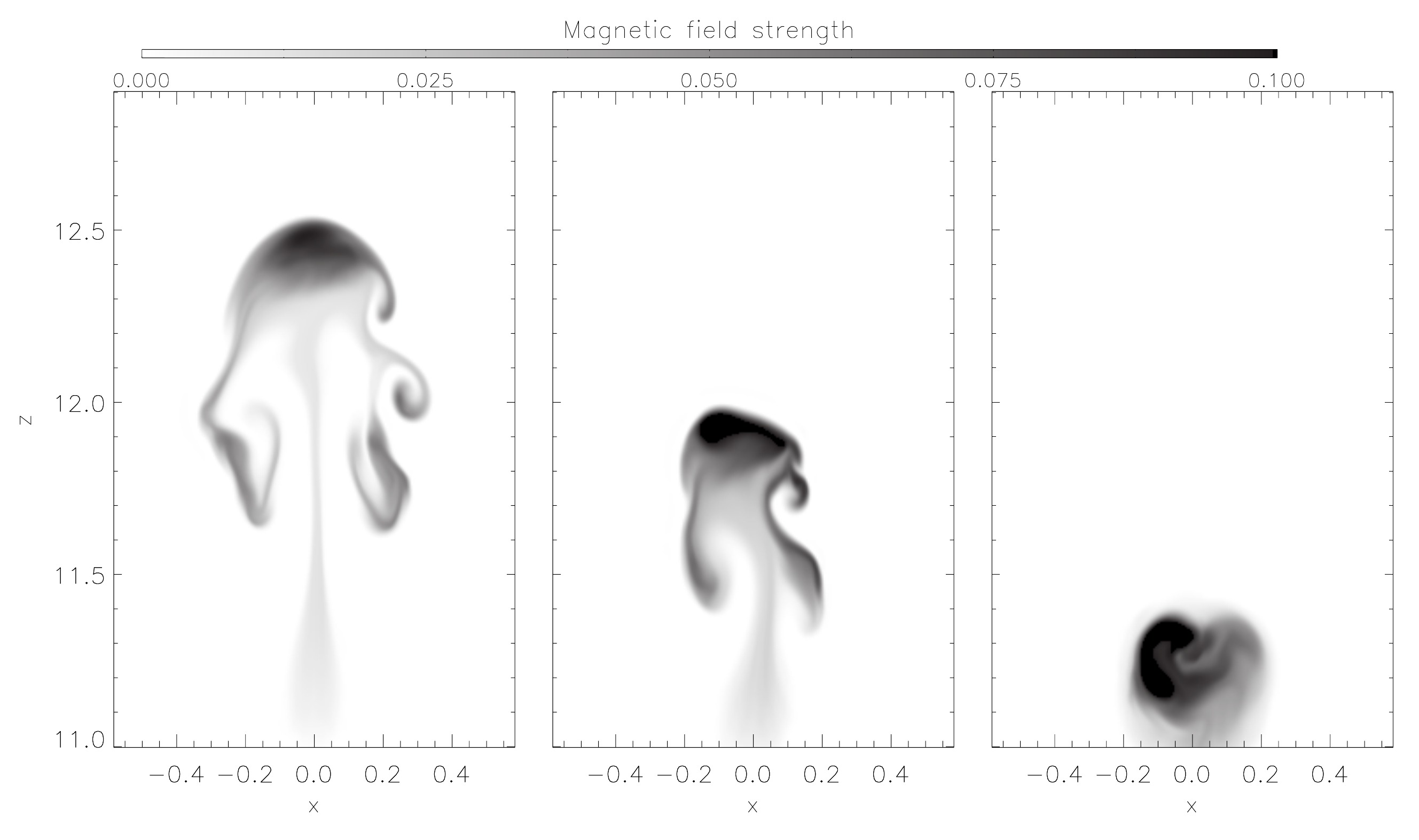}
\caption{\label{fig:bz3x} {\footnotesize A series of snapshots of the 
cross section of the longitudinal magnetic field are shown for the run
$\chi_{1.5}^{}\lambda_{1/8}^{}H$ at $t=65.95$, 
at $y=(-0.762, -2.25,-2.25)$, from left to right, respectively.}}
\end{figure*}

The vorticity changes associated with the shedding of vortex rolls from the
wake of the rising tube [item (\ref{enum:ii}) in the list right before
Section~\ref{sec:inhomogeneity}], and the corresponding changes in the 
lift acceleration also cause discernible perturbations in the motion of the
tube. Figure~\ref{fig:lift2} shows the trajectory of the tube axis on the 
transverse $xz$ plane at fixed distances from the apex for various
experiments with the highest AMR levels but with different values of twist
and $\Omega$-loop curvature. The time interval shown is $t= 0$ to $t\sim70$. 
All trajectories shown in the figure have an oscillating
shape: the oscillation corresponds to the shedding of vortex rolls into the
Von Karman street. The three solid lines, in particular, correspond to the
three cases $\chi_{1.5}^{}\lambda_{1/4}^{}H$ (red),
$\chi_{1.5}^{}\lambda_{1/8}^{}H$ (blue) and $\chi_{1.5}^{}\lambda_{1/16}^{}H$
(black), with the trajectories calculated on the vertical plane $y = 3.1$; we
have provided velocity and field strength maps at the apex for those cases in
Figure~\ref{fig:bz3b} (top-left, top-right and bottom-left panels,
respectively). The total vorticity of the vortex rolls in the wake of those
tubes increases for decreasing twist. Correspondingly, when the wake becomes
asymmetric and vortex rolls are alternatively shed from either side of the
tube into the Von Karman street, the associated elongation of the
oscillations in the solid curves in Figure~\ref{fig:lift2} becomes larger: this
effect is quite marked for the low-twist case (black solid line) and
marginally apparent for the red and blue curves.  The actual asymmetry of the
wake for the intermediate-twist case ($\chi_{1.5}^{}\lambda_{1/8}^{}H$, i.e.,
the case for the blue solid curve in Figure~\ref{fig:lift2}) can be seen in
Figure~\ref{fig:bz3x}, with views provided for vertical cuts at the same
instant ($t=65.95$) and three different distances from the tube apex, namely,
$y=-0.762$, $y=-2.25$ and $y=-3.74$; for comparison, one can also check the
3D volume rendering of the vorticity for this tube shown in
Figure~\ref{fig:vrtz3d}. Another effect apparent in the three solid curves of
Figure~\ref{fig:lift2} is that the global inclination of the trajectory is
very small for the black curve and larger but rather similar for the red and
blue curves. The global inclination of the trajectory is due to the effect
discussed in the previous subsection~\ref{sec:inhomogeneity};
correspondingly, when the twist is low (as for the case of the black solid
curve in the figure) the vorticity generation due to this effect is also less
important, and the motion away from the initial vertical plane less
marked. 

Still in Figure~\ref{fig:lift2}, we now compare the three curves in red,
which explore the dependence with the curvature of the $\Omega$-loop by
choosing $\chi=2.1$ ($\chi_{2.1}^{}\lambda_{1/4}^{}H$, dashed
line), $\chi=1.5$ ($\chi_{1.5}^{}\lambda_{1/4}^{}H$, solid) and
$\chi=1$ ($\chi_{1}^{}\lambda_{1/4}^{}H$, dash-dotted) on the
planes $y=3.7$, $y = 3.1$, and $y = 2$, respectively. The choice of different
$y$ values for those cases is because of the smaller longitudinal size of the
$\Omega$-loop for larger curvature cases. 
We see that there are only minor
differences in the oscillations and global inclination of those curves. This
is in line with the results of Sections~\ref{sec:frag} and \ref{sec:highre}
where we saw that in the high-AMR experiments the cross sectional structure
of the head of the tube and the amount of magnetic flux it retains is similar
irrespective of the value of $\chi$. Here we show that also the net vorticity
generation and the vorticity loss to the wake do not seem to depend on the
curvature to any important extent.

\section{Discussion and conclusions}
\label{sec:conc}

We have carried out a set of 2D and 3D MHD simulations of buoyant magnetic
$\Omega$-loops rising through a stratified layer covering a range of values
of the loop curvature (as measured through the $\chi$ parameter, with $\chi
\in [1,\infty]$), field line twist ($\lambda \in [0,1/4]$) and effective
Reynolds number. We have reached comparatively high Reynolds numbers in the 3D
experiments by using the Adaptive Mesh Refinement (AMR) 
algorithm of the Flash code. From the present study, we argue that the
evolution of $\Omega$-loops can be quite different when we compare the
high-diffusion cases [with $Re \lesssim \hbox{O}(10)$, say] to cases with $Re
\sim \hbox{O}(100)$. In the more diffusive cases with up to a moderate level of twist 
($\lambda=1/8$), the rising tube does not
evolve into the simple configuration of a rising compact head with a trailing
wake. In those 
cases the evolution strongly depends on the curvature of the tube axis; the
dependence is substantially diminished in the high-resolution,
low-diffusivity cases. A clear example has been shown in the text
(Figures~\ref{fig:flucur}, \ref{fig:bz3a}, and
\ref{fig:bz3a_low}).  In the low AMR regime (high diffusion, Figure~\ref{fig:bz3a_low})  
without curvature, the tube's head splits into two big
counterrotating vortices which drift apart due to the lift force when the
rising tube is horizontal; instead, in the case of an $\Omega$-loop with high
enough curvature, the vortices are kept in the neighborhood of the loop axis.
This effect comes about through the gas pressure deficit created near the
loop apex by the plasma expansion in the direction of the $\Omega$-loop axis
whereas the plasma does not expand in that direction when the tube rises
keeping a straight, horizontal shape. That there is a large dependence on
curvature in those of our experiments without AMR is concomitant with the
results of ~\citet{abbett2000}, who found that a sufficient level of
$\Omega$-loop curvature can help maintain the coherence of the tube. In
contrast, in the high AMR regime (low diffusion) for the same value of
$\lambda$ (Figure~\ref{fig:bz3a}), the vorticity is concentrated in a thin
boundary layer, no splitting of the tube's head into vortices is seen, and
there are no important differences in the results of the experiment in
dependence of the loop curvature.

In parallel to the foregoing, as part of our parametric study we have
considered the dependence of the fragmentation of the tube with the magnetic
field twist, focusing on cases with high AMR and a fixed curvature value
(Figures~\ref{fig:flutw} and \ref{fig:bz3b}). The amount of magnetic flux
retained in the head of the tube as a function of field line twist in our 3D
simulations follows a pattern as reported in 2D studies in the past
\citep{moreno1996,emonet1998,mark2006}: as the twist increases and as a
result of the magnetic field tension, the tube's head is more rounded and
loses less magnetic flux to the wake. For $\lambda \gtrsim 1/8$ the head of
the tube retains in all cases in excess of $\approx 50\%$ of the original
magnetic flux (more than $70\%$ for $\lambda = 1/4$). The cross section of
the apex of untwisted tubes, in turn, develops into a bar-shaped structure
along the rise. This bar is a result of a horizontal expansion that is not
counteracted by magnetic tension and is observed in all simulations without
twist regardless of the curvature of the axis and of the Reynolds number. 

Another finding of this work concerns the flanks of those $\Omega$-loops 
with sufficiently twisted field lines to rise keeping a coherent tube-like 
structure: we see the flanks move away from their initial vertical plane 
in an antisymmetric fashion. We argue that such
antisymmetric horizontal drift comes about because those tubes develop a
global vortical motion around the tube axis in each flank, with vorticity
pointing toward the apex of the $\Omega$-loop: the ensuing lift force then
leads to the antisymmetric motion of the flank away from the initial vertical
plane. The vorticity is generated as a result of the
inhomogeneous distribution of field line pitch on magnetic flux surfaces
as one slides along the loop downward from the apex. 
This inhomogeneous distribution of field line pitch is directly caused by the
expansion of the rising tube segments and leads to an unbalanced Lorentz
force in the azimuthal direction around the tube axis, i.e., to a torque that
makes the tube rotate around its axis, with associated generation of
vorticity. \citet{Fan:2008qf} found that, when seen from above, the head of
the rising $\Omega$-loops in her experiment was increasingly
tilted with respect to the direction of the tube in its original vertical
plane. Considering the Maxwell stresses exerted on the end cross sections of
a tube segment, she explained the tilt through the lack of compensation of
the torque of those stresses when the segment is at or near the apex of the
loop. Instead, no tilt is produced for a straight tube. We believe the
physical process leading to the antisymmetric behavior of the flanks in our
rising $\Omega$-loop to be different to the one described by
\citet{Fan:2008qf}. In fact, the mechanism explained in Section
\ref{sec:inhomogeneity} and summarized just above does operate also for
straight tubes with varying cross-sectional radius as one slides along the
tube axis. Also, the motion of the flanks in our case is an indirect effect
of the Lorentz force: the latter generates vorticity in the tube which then
leads, via the lift force, to the motion off the vertical plane.

The motion of the flanks off their initial vertical plane is accompanied by
an oscillatory pattern (Figures~\ref{fig:lift1} and ~\ref{fig:lift2}). The
oscillation is due to the shedding of vortex rolls to the wake and the
corresponding creation of a Von Karman vortex street. We find that the
amplitude of the oscillation increases as one considers less twisted tubes,
because the rolls shed to the wake are larger and carry more vorticity the
less twisted the initial tube. In contrast, there is no important dependence
of the amplitude of oscillation with the curvature of the head of the
$\Omega$-loop. In the papers by \citet{emonet2001, mark2006} the formation of
a Von Karman street had been reached by imposing a net vorticity inside the
tube at the beginning of the experiments. In the present case, the net
vorticity is generated self-consistently through the mechanism indicated in
the previous paragraph and in Section~\ref{sec:inhomogeneity} and the vortex
shedding thus follows naturally. The 2D simulations of rising buoyant tubes
by \citet{Hughes:1998sf} had an adaptive grid refinement algorithm and
reached high Reynolds numbers, like, e.g., $4\,10^3$. They showed the
trajectory of the axis of the rising tube in a vertical plane and compared it
with that obtained in an experiment with a higher level of diffusion: the
former showed what can be described as epicycles superimposed on a smooth
trajectory (see their Figure 3). From the figures in their paper one gathers
that vortex shedding was taking place in their experiment. \citet{emonet2001}
showed that trajectories for tubes with a high level of vorticity (more
precisely, for a high value of the ratio defined in
Equation~\ref{eq:param_emonet_etal_2001}) could show epicycles and be far from
simple, smooth sinusoidal paths. In three dimensions it is difficult to reach
Reynolds numbers of a few thousand. On the other hand, as shown in the
present paper, the extrapolation of 2D results to three dimensions is not
straightforward since, among other things, the sideways motion of the tube
can be countered by the Lorentz force associated with the tension of the
longitudinal field component. The final answer to the evolution of 3D tubes
must await the use of codes that allow experiments with very low diffusivity.

A natural extension of the current work would be to the
case of magnetic ropes rising across a spherical shell mimicking the solar
convection zone, i.e., at least using an unperturbed stratification closer to
that of the actual solar convection zone and spherical
geometry. Attempts in this direction have been made by \citet{Fan:2008qf} or,
including self-consistent large-scale convection flows, by
\citet{Jouve:2009bh, Jouve:2013kx, Pinto:2013uq}. As numerical methods and
computing facilities improve, it will be possible to reach Reynolds numbers
in that kind of experiment well above $100$, in which the effects shown in
this paper should become apparent.  In fact, the phenomenon of flux tube
splitting, and the influence thereon of the field line twist and curvature of
the loop axis, could be preserved also in the presence of convection flows
since they are essentially associated with the internal flux tube dynamics
(barring effects associated to regions of strong shear in the convective
cells). In any case, the dynamics of magnetic tubes moving in 3D environments 
is complicated enough to have to await the actual numerical experiments to decide 
on the approximate validity or otherwise of the simplifying assumptions made in 
this paper.

\section{Acknowledgements}

The authors acknowledge the support by the Spanish Ministry of Economy 
and Competitiveness through projects AYA2011-24808 and AYA2014-55078-P 
and by the European Commission through the SOLAIRE Network (MTRN- CT-2006-035484). 
JMS is supported by NASA grants NNX11AN98G, NNM12AB40P and NASA 
contracts NNM07AA01C (Hinode), NNG09FA40C (IRIS), and NNX14AI14G (HGCR grant). We gratefully 
acknowledge the resources, technical expertise and assistance provided at the 
MareNostrum (BSC/CNS, Spain) and LaPalma (IAC, Spain) supercomputer installations, 
Notur project, and the Pleiades cluster through the computing project s1061 from the High 
End Computing (HEC) division of NASA as well as the computer and supercomputer 
resources of the Research Council of Norway through grant 170935/V30 and through 
grants of computing time from the Programme for Supercomputing. The FLASH computer 
code used in this work was in part developed by the DOE-supported ASC/Alliance Center 
for Astrophysical Thermonuclear Flashes at the University of Chicago. For 3D visualization 
UCAR's VAPOR package (\citealt{2007NJPh....9..301C};
http://www.vapor.ucar.edu) and Tecplot software were used. Thanks are also due to S. Guglielmino for help and 
discussions during this project and to M. Jin for his help with the 3D 
visualization of Figure~\ref{fig:3db} for the online version of the paper.

\appendix

\section{The determination of the effective Reynolds number}~\label{sec:rey}

Determining the effective Reynolds number of a simulation is not simple. For
multidimensional experiments of rising flux tubes, the problem is alleviated
by using the theory of boundary layer formation in flows around
straight cylinders, which predicts a layer width of order $D\,Re^{-1/2}$,
with $D$ the diameter of the cylinder. How well this law applies also to
rising magnetic tubes in a stratified medium was studied in some detail by
\citet{emonet1998}.  We follow those results in a similar way as
\citet{mark2006} and assume that
\begin{equation}\label{eq:reynolds_calibrated}
Re=K\left(\frac{D}{L_b}\right)^2,
\end{equation}

\noindent where $D$ here stands for the transverse size of the tube at a
given time, $L_b$ for the width of the (magnetic or viscous) boundary layer
around it, and $K$ is a calibration factor of order unity introduced in this
paper and explained below. In order to calculate $D$, we first calculate the
area $A$ of the tube cross section by determining the domain where $B > \Beq$, with
$\Beq$ given through Equation~(\ref{eq:equipartition}), and then specify $D =2
\sqrt{A/\pi}$. This prescription works well for the simulations with medium
and high AMR; for those in the low AMR regime the cross sectional area $A$
is obtained through the condition $B > 0.25 \Beq$. We have chosen this 
criteria because the main body of the tube for simulations with high diffusion and 
enough curvature has two vortex rolls with considerably large velocities and still 
maintains the magnetic structure of the tube's head during their evolution 
(see Section~\ref{sec:lowre}).  The last column in 
Table~\ref{tab:sim2} shows the diameter calculated in this way for each
simulation at time $t\sim50$. The boundary layer width, $L_b$, can be
measured in different ways \citep{emonet1998}. In a reference system moving
upward with the speed of the center of the tube apex, the external medium is
seen flowing downward around the head. The transition between the velocity of the
external medium and the tube interior occurs in an abrupt way precisely in
the boundary layer. The best determination of the boundary layer width can be
done by measuring $v_x$ along a vertical axis slightly offset
by one grid point from the mid vertical axis of the tube head: $L_b$ is determined 
as the width of the layer in which $v_x$ 
decreases from the external value to $1/e$ of it. The value obtained in this
way for $L_b$ typically spans about six grid cells for any level of AMR.

The seventh column in Table~\ref{tab:sim2} lists the Reynolds number for
each simulation at time $t\sim 50$ using
Equation~\ref{eq:reynolds_calibrated} with $K$ defined so as 
to allow comparison with the numbers given by \citet{mark2006}. To that end, we note
that the Reynolds number varies in time mainly because the 
tube diameter increases along the rise due to the decrease of the external
pressure. The simulations by \citet{mark2006} reached larger Reynolds numbers
than ours for the same spatial resolution because they calculated the
Reynolds number at substantially later instants: their $2.5$D simulations allowed them to
include a domain spanning a larger number of pressure scale-heights than in
the present paper.  Therefore, in order to facilitate the comparison between
the two studies, we decided to calibrate our estimates of $Re$ to the values
quoted by \citet{mark2006}. We choose a $2.5$D simulation in our set for
which the refinement level is the same in both studies. Comparing the value
of $Re$ obtained by us and the number quoted by \citet{mark2006} at a later
time, we find a factor $1.75$ between them. We then multiply the values of
$Re$ in our experiments with that factor, i.e., we set $K=1.75$ in
Equation~\ref{eq:reynolds_calibrated}. This recalibration 
is justified for our horizontal tube (i.e., 2.5D) experiments for the reasons
mentioned, but, of course, only leads to a rough estimate of the Reynolds number at
later times for the general 3D cases. 

In our simulations, the viscous and magnetic boundary layers to
have similar thickness, about 6 grid cells, indicating that $Re \approx Re_m$. 
This is because the viscous and magnetic diffusion stem from the numerical
truncation inherent in the scheme. 

While the maximum refinement level has a direct impact on $Re$, it is not the 
only relevant factor. Other properties of the tube, for instance, the twist and 
the curvature, affect the evolution of the tube and thus play indirect roles 
on the effective Reynolds number achieved by the simulation. In fact, the twist 
and the curvature can change the size of the tube's head such as is described in 
more detail in our results (see as well the sixth, seventh and tenth columns in 
Table~\ref{tab:sim2}).

\bibliographystyle{aa}

\end{document}